\documentclass[11pt,a4paper]{article}
\usepackage[a4paper,total={6.00in, 9.25in}]{geometry}
\usepackage{graphicx, hyperref, enumitem, xcolor}
\usepackage[round]{natbib}
\usepackage{setspace} 
\usepackage[normalem]{ulem} 
\usepackage{amsmath,amssymb}
\usepackage{amsthm}
\theoremstyle{plain}

\newtheorem{definition}{Definition}

\title{\textbf{Investigation of finite-sample properties of robust location and scale estimators}}
\author{
Chanseok Park
~and Haewon Kim\\
Applied Statistics Laboratory\\
Department of Industrial Engineering\\ Pusan National University\\
Busan 46241, Korea
\and
Min Wang\thanks{Corresponding author. Email: min.wang3@utsa.edu} \\
Department of Management Science and Statistics \\
The University of Texas at San Antonio \\
San Antonio, TX, USA
}
\date{}

\begin{document}
\maketitle
\begin{abstract}
When the experimental data set is contaminated,
we usually employ robust alternatives to common location and scale estimators
such as the sample median and Hodges-Lehmann estimators for location and the sample median absolute deviation and Shamos estimators for scale.
It is well known that these estimators have high positive asymptotic breakdown points
and are Fisher-consistent as the sample size tends to infinity.
To the best of our knowledge, the finite-sample properties of these estimators,
depending on the sample size, have not well been studied in the literature.

In this paper, we fill this gap by providing their closed-form finite-sample breakdown points
and calculating the unbiasing factors and
relative efficiencies of the robust estimators through the extensive Monte Carlo simulations up to the sample size 100.
The numerical study shows that the unbiasing factor
improves the finite-sample performance significantly.
In addition, we provide the predicted values for the unbiasing
factors obtained by using the least squares method which can be used
for the case of sample size more than 100.

\smallskip
\noindent\textbf{\uppercase{Keywords}}: breakdown, unbiasedness,
robustness, relative efficiency
\end{abstract}

\section{Introduction}
Estimation of the location and scale parameters of a distribution, such as the mean and standard deviation of a normal population,
is a common and important problem in the various branches of science and engineering including:
biomedical, chemical, materials, mechanical and industrial engineering, etc.
The quality of the data plays an important role in estimating these parameters.
In practice, the experimental data are often contaminated
due to the measurement errors, the volatile operating conditions, etc. 
Thus, robust estimations are advocated as alternatives to commonly used location and scale estimators 
(e.g., the sample mean and sample standard deviation) for estimating these parameters.
For example, for the case where some of the observations are contaminated
by outliers, we usually adopt the sample median and
Hodges-Lehmann \citep{Hodges/Lehmann:1963} estimators for location and the sample median absolute deviation \citep{Hampel:1974} and Shamos \citep{Shamos:1976} estimators for scale, because these estimators have a large breakdown point and perform well in either the presence or absence of outliers.
	
The breakdown point is a common criterion for measuring the robustness of an estimator.
The larger the breakdown point of an estimator, the more robust it is.
The finite-sample breakdown point based on the idea of \cite{Hodges:1967} is defined as the maximum proportion of incorrect or arbitrarily observations
that an estimator can deal with without making an egregiously incorrect value.
The breakdown point of an estimator is a measure of its resistance to data contamination \citep{Wilcox:2016}.
For example, the breakdown points of the sample mean and the sample median are 0 and $1/2$,
respectively. Given that the breakdown point can generally be written as a function of the sample size, 
we provide the finite-sample breakdown points for the various location and scale estimators mentioned above. 
It is shown that when the sample sizes are small, they are noticeably different from the asymptotic breakdown point,
which is the limit of the finite-sample breakdown point when the sample size tends to infinity.
	
It deserves mentioning that for a robust scale estimation, the sample median absolute deviation (MAD)
and Shamos estimators not only have positive asymptotic breakdown points,
but also are Fisher-consistent \citep{Fisher:1922} as the sample size goes to infinity.
However, when the sample size is small, they have serious biases and thus provide inappropriate estimation.
Some bias-correction techniques are commonly adopted to improve the finite-sample performance of these estimators.
For instance, \cite{Williams:2011} studied the finite-sample correction factors through computer simulations for several simple robust estimators of the standard deviation of a normal population, which include the MAD, interquartile range, shortest half interval, and median moving range.
Later on, under the normal distribution, \cite{Hayes:2014} obtained the finite-sample bias-correction factors
of the MAD for scale. They have shown that finite-sample bias-correction factors can significantly
eliminate systematic biases of these robust estimators, especially when the sample size is small.

To the best of our knowledge,
finite-sample properties of the MAD and Shamos estimators have received little attention in the literature except for some references covering topics for small sample sizes. This observation motivates us to employ the extensive Monte Carlo simulations to obtain the empirical biases of these estimators. Given that the empirical variance of an estimator is one of the important metrics for evaluating an estimator, we also obtain the values of the finite-sample variances of the median, Hodges-Lehmann (HL),
MAD and Shamos estimators under the standard normal distribution, which are not fully provided in the statistics literature.
Numerical results show that the unbiasing factor improves
the finite-sample performance of the estimator significantly.
In addition, we provide the predicted values
for the unbiasing factors obtained
by the least squares method which can be
used for the case of the sample size more than 100.

The rest of this paper is organized as follows. In Section \ref{SEC:breakdown}, we derive the finite-sample breakdown points for robust location estimators and robust scale estimators, respectively. Through the extensive Monte Carlo simulations,
we calculate the empirical biases of the MAD and Shamos estimators in Section \ref{SEC:biase}
and the finite-sample variances of the median, HL, MAD, and Shamos estimators in Section \ref{SEC:empirical}.
An example is provided to illustrate the application of the proposed methods
to statistical quality control charts for averages and standard deviations in Section \ref{section:Example}.
Some concluding remarks are given in Section \ref{section:Concluding}.
	
\section{Finite-sample breakdown point} \label{SEC:breakdown}
In this section,
we introduce the definitions of the bias and breakdown of an estimator $\hat{\theta}$
using Definition 1.6.1 of \cite{Hettmansperger/McKean:2010}
and Section~2 of \cite{Hampel/etc:1986}, which is based on the idea of \cite{Hodges:1967}.
According to these definitions,
we derive the explicit finite-sample breakdown points of the robust
estimators for location and scale parameters.
\begin{definition}[Bias] \label{DEF:bias}
Let $\mathbf{x}=\{x_1,x_2,\ldots,x_n\}$ be a realization of a sample
$\mathbf{X} = \{X_1,X_2,\ldots,X_n\}$.
We denote the corrupted sample of any $m$ of the $n$ observations as
\[
\mathbf{x}_{(m)}^* = \{x_1^*,x_2^*,\ldots,x_m^*, x_{m+1}, \ldots, x_n\}.
\]
The \emph{bias} of an estimator $\hat{\theta}$ is defined as
\[
\mathrm{Bias}(\hat{\theta} ; \mathbf{x}_{(m)}^*, \mathbf{x})
= \sup| \hat{\theta}(\mathbf{x}_{(m)}^*) - \hat{\theta}(\mathbf{x}) |,
\]
where the $\sup$ is taken over all possible corrected observations $x_i^*$
with $x_i$ being fixed at their original observations in the realization of $\mathbf{x}_{(m)}^*$.
\end{definition}
If $\mathrm{Bias}(\hat{\theta} ; \mathbf{x}_{(m)}^*, \mathbf{x})$
makes an arbitrarily large value (say, $\infty$),
we say that the estimator $\hat{\theta}$ has broken down.
\begin{definition}[Breakdown] \label{DEF:breakdown}
We define the finite-sample breakdown as
\begin{equation} \label{EQ:breakdown1}
\epsilon_n =
\frac{1}{n}\max\big\{m:\mathrm{Bias}(\hat{\theta};\mathbf{x}_{(m)}^*,\mathbf{x})<\infty \big\}.
\end{equation}
\end{definition}
This approach to breakdown is called $\epsilon$-replacement
breakdown because observations are replaced by corrupted values.
For more details, readers are referred to \citep{Donoho/Huber:1983},
Subsection 2.2a of \cite{Hampel/etc:1986}, and Subsection 1.6.1
of \cite{Hettmansperger/McKean:2010}. The sample median and HL estimators for robust location estimation
are considered in Subsection~\ref{SEC:breakdown-location} and
the MAD and Shamos estimators for robust scale estimation are
in~Subsection \ref{SEC:breakdown-scale}.

\subsection{Robust location estimators} \label{SEC:breakdown-location}
It is well known that the asymptotic breakdown point of the sample median
is 1/2; see, for example, Example~2 in Section 2.2 of  \cite{Hampel/etc:1986}.
For the HL estimator, the asymptotic breakdown point
is given by $1-1/\sqrt{2}\approx0.293$; see Example~8 of Section 2.3 of \cite{Hampel/etc:1986} for more details.
Note that the sample median and HL estimators are in closed-forms
and are location-equivariant in the sense that
\[
\hat{\theta}(X_1+b,X_2+b,\ldots,X_n+b) = \hat{\theta}(X_1,X_2,\ldots,X_n)+b.
\]
However, in many cases,
the finite-sample breakdown points can be noticeably different from the asymptotic breakdown point, especially when the sample size $n$ is small.
For instance, when $n = 10$, we observe from Equation (\ref{EQ:e-median}) that the finite-sample breakdown point for the median is 0.4, which is different from its asymptotic breakdown point of 0.5.
This motivates the need for further investigating the finite-sample properties of these robust estimators.

Suppose that we have a sample, $\{X_1, X_2, \ldots, X_n\}$.
If the sample size $n$ is given by an odd number (say, $n=2k+1$),
the median can be resistant to $k$ corrupted observations.
Thus, the finite-sample breakdown point of the median with $n=2k+1$ is given by
$\epsilon_n = k/n$.
Next, if the sample size $n$ is given by an even number (say, $n=2k$),
then the median can be resistant to $k-1$ corrupted observations which gives
$\epsilon_n = (k-1)/n$.
Let $\lfloor\cdot\rfloor$ be the floor function. That is,
$\lfloor x \rfloor$ is the largest integer not exceeding $x$.
Then we have $\lfloor (n-1)/2 \rfloor = k$ for $n=2k+1$
and $\lfloor (n-1)/2 \rfloor = k-1$ for $n=2k$ so that the finite-sample breakdown point
can be written by the single formula
$\lfloor (n-1)/2 \rfloor / n$ for both odd and even cases.
The finite-sample breakdown point of the median is thus given by
\begin{equation} \label{EQ:e-median}
\epsilon_n = \frac{\lfloor (n-1)/2 \rfloor}{n}.
\end{equation}

It should be noted that Equation (1.6.1) of \cite{Hettmansperger/McKean:2010} defines
the finite-sample breakdown point to be
\(
\epsilon_n^\star =
\min\big\{m/n:\mathrm{Bias}(\hat{\theta};\mathbf{x}_{(m)}^*,\mathbf{x})=\infty \big\}.
\)
According to this definition, the finite-sample breakdown point of the median is
given by $\epsilon_n^\star = {\lfloor (n+1)/2 \rfloor}/{n}$
which is always greater than $1/2$ for any odd value of $n$.
For example, $\epsilon_n^\star = 2/3$ with $n=3$.
Also, for the sample mean, $\epsilon_n^\star$ is always $1/n$, not zero.
For example,  $\epsilon_n^\star=1/2$ with $n=2$ although its asymptotic breakdown point is zero.
On the other hand, based on the finite-sample breakdown point in Equation (\ref{EQ:breakdown1}),
we have $\epsilon_n = 1/3$ with $n=3$ for the median
and $\epsilon_n=0$ always for the sample mean.
It deserves mentioning that there are several variants of the breakdown point. We refer the interested readers to Remark~1 in Section 2.2 of \cite{Hampel/etc:1986}
and Chapter~11 of \cite{Huber/Ronchetti:2009} for more details.
In this paper, we follow Equation (\ref{EQ:breakdown1}) in Definition~\ref{DEF:breakdown} to define the finite-sample breakdown.


Using the fact that $\lfloor x \rfloor$ can be rewritten
as $\lfloor x \rfloor = x - \delta$ where $0 \le \delta < 1$, we have
\[
\epsilon_n = \frac{\lfloor (n-1)/2 \rfloor}{n}
= \frac{1}{2} - \frac{1}{2n} - \frac{\delta}{n}.
\]
Thus, the asymptotic breakdown point of the median is obtained by taking the limit of
the finite-sample breakdown point as $n\to\infty$, which provides that $\epsilon = \lim_{n\to\infty} \epsilon_n = 1/2$.

The HL estimator \citep{Hodges/Lehmann:1963} is defined as the median of all pairwise averages of the sample observations and is given by
\[
\mathop{\mathrm{median}} \Big( \frac{X_i+X_j}{2} \Big).
\]
Note that the median of all pairwise averages can be calculated for the three cases:
(i) $i<j$, (ii) $i\le j$, and  (iii) $\forall(i,j)$, where $i,j = 1,2,\ldots,n$.
We denoted these three versions as
\[
\mathrm{HL1} = \mathop{\mathrm{median}}_{i<j} \Big( \frac{X_i+X_j}{2} \Big), \quad
\mathrm{HL2} = \mathop{\mathrm{median}}_{i\le j} \Big( \frac{X_i+X_j}{2} \Big), \quad
\mathrm{HL3} = \mathop{\mathrm{median}}_{\forall(i,j)} \Big( \frac{X_i+X_j}{2} \Big),
\]
respectively.
Note that each of the paired averages $(X_i+X_j)/2$
where $i \le j$ and $i,j = 1,2,\ldots,n$ is called a Walsh average
\citep{Walsh:1949}.
In what follows, we first derive the finite-sample breakdown point
for the $\mathrm{HL3}$ and then use a similar approach to derive
the finite-sample breakdown points for $\mathrm{HL1}$ and $\mathrm{HL2}$.
It is noteworthy that all three versions are \emph{asymptotically} equivalent
as mentioned in Example 3.7 of \cite{Huber/Ronchetti:2009} although
they are practically different with a sample of \emph{finite} size.

The basic idea of how to derive the finite-sample breakdown point
for the HL estimators was proposed by \cite{Ouyang/Park/etc:2019},
but they did not provide explicit formulae.
In this paper, we provide the method of how to derive the explicit formulae.
Suppose that we make $k$ of the $n$ observations arbitrarily large,
where $k=0,1,2,\ldots,n$.
Notice that there are $n\times n$ paired average terms in the HL3 estimator: $(X_i+X_j)/2$,
where $i,j = 1,2,\ldots,n$.
Because the HL3 estimator is the median of the  $n\times n$ values,
the finite-sample breakdown point cannot be greater than
$\lfloor (n^2-1)/2\rfloor/n^2$ due to Equation (\ref{EQ:e-median}).
If we make $k$ of the $n$ observations arbitrarily large,
then the number of arbitrarily large paired averages
becomes $n^2-(n-k)^2$. These two facts provide the following relationship
\[
\frac{n^2-(n-k)^2}{n^2} \le \frac{\lfloor (n^2-1)/2\rfloor}{n^2},
\]
which is equivalent to $k^2 - 2nk + \lfloor (n^2-1)/2\rfloor \ge 0$.
The finite-sample breakdown point of the  $\mathrm{HL3}$ is then given by
$\epsilon_n = k^{*} / n$, where
\begin{equation} \label{EQ:kstar}
k^{*} = \max\Big\{k: k^2-2nk+\lfloor(n^2-1)/2\rfloor\ge0
\textrm{~and~} k=0,1,2,\ldots,n \Big\}.
\end{equation}
To obtain an explicit formula for Equation (\ref{EQ:kstar}), we let $f(x) = x^2 - 2nx + \lfloor (n^2-1)/2\rfloor$.
Since $f'(x)=2(x-n)$, $f(x)$ is decreasing for $x<n$.
The roots of $f(x)=0$ are given by
\[
x_1 = n-\sqrt{n^2 - \lfloor (n^2-1)/2\rfloor} \textrm{~~~and~~~}
x_2 = n+\sqrt{n^2 - \lfloor (n^2-1)/2\rfloor}.
\]
Since $k$ is an integer and $k \le n$, we have $k^*=\lfloor x_1 \rfloor$,
that is, $k^*  = \Big\lfloor n - \sqrt{n^2-\lfloor (n^2-1)/2\rfloor} \Big\rfloor.$
Then we have the closed-form finite-sample breakdown point of the $\mathrm{HL3}$
\begin{equation} \label{EQ:e-HL3}
\epsilon_n = \frac{\Big\lfloor n-\sqrt{n^2-\lfloor(n^2-1)/2\rfloor}\Big\rfloor}{n}.
\end{equation}
The asymptotic breakdown point of $\mathrm{HL3}$ is given by
$\epsilon = \lim_{n\to\infty} \epsilon_n$.
Using $\lfloor x \rfloor = x-\delta$ where $0\le\delta<1$,
we can rewrite Equation (\ref{EQ:e-HL3}) as
\[
\epsilon_n = \frac{n-\sqrt{n^2-  (n^2-1)/2 +\delta_1}-\delta_2}{n},
\]
where $0\le \delta_1 <1$ and $0\le \delta_2 <1$.
Thus, we have $\epsilon = 1-1/\sqrt{2} \approx 29.3\%$.

In the case of the $\mathrm{HL1}$ estimator, there are
$n(n-1)/2$ Walsh averages. Since the $\mathrm{HL1}$ estimator is the median
of the $n(n-1)/2$ Walsh averages, the finite-sample breakdown point cannot be greater than
$\lfloor\{n(n-1)/2-1\}/2 \rfloor/ \{n(n-1)/2\}$
due to Equation (\ref{EQ:e-median}) again.
If we make $k$ observations arbitrarily large with $k\le n-1$,
then there are $n(n-1)/2 - (n-k)(n-k-1)/2$ arbitrarily large Walsh averages.
Thus, the following inequality holds
\[
\frac{n(n-1)/2 - (n-k)(n-k-1)/2}{ n(n-1)/2}
\le \frac{1}{n(n-1)/2} \Big\lfloor \frac{n(n-1)/2-1}{2} \Big\rfloor,
\]
which is equivalent to $k^2 - (2n-1)k + 2\lfloor(n^2-n-2)/4\rfloor \ge 0.$
In a similar way as done for Equation (\ref{EQ:kstar}), we let $k^*$ be the largest integer $k$ satisfying
the above, where $k=0,1,2,\ldots,n-1$.
For convenience, we let $f(x)=x^2 - (2n-1)x + 2\lfloor(n^2-n-2)/4\rfloor$.
Then $f(x)$ is decreasing for $x<n-1/2$ due to $f'(x)=2x-(2n-1)$ and
the roots of $f(x)=0$ are given by
$n-1/2 \pm \sqrt{(n-1/2)^2-2\lfloor(n^2-n-2)/4\rfloor}$.
Thus, using the similar argument to that used for the $\mathrm{HL3}$ case, we have
$k^* = \Big\lfloor n-1/2 - \sqrt{(n-1/2)^2-2\lfloor(n^2-n-2)/4\rfloor} \Big\rfloor.$
Then we have the closed-form finite-sample breakdown point of the $\mathrm{HL1}$
\begin{equation} \label{EQ:e-HL1}
\epsilon_n
= \frac{\Big\lfloor n-1/2 - \sqrt{(n-1/2)^2-2\lfloor(n^2-n-2)/4\rfloor}\Big\rfloor}{n}.
\end{equation}
It should be noted that we also have
$\epsilon=\lim_{n\to\infty}\epsilon_n=1-1/\sqrt{2}$.

Similar to the case of the $\mathrm{HL1}$,
we obtain that the closed-form finite-sample breakdown point of the $\mathrm{HL2}$ estimator is given by
\begin{equation} \label{EQ:e-HL2}
\epsilon_n
= \frac{\Big\lfloor n+1/2 - \sqrt{(n+1/2)^2-2\lfloor(n^2+n-2)/4\rfloor}\Big\rfloor}{n}.
\end{equation}

In Table~\ref{TBL:breakdown}, we provide the finite-sample breakdown points of
the three HL estimators along with the median, the MAD and the Shamos estimators, which
will be covered in the following subsection.
Also we provide the plot of these values in Figure~\ref{FIG:breakdown}.
By comparing these values, we observe that
\[
\epsilon_n(\textrm{HL2}) \ge \epsilon_n(\textrm{HL3}) \ge \epsilon_n(\textrm{HL1}).
\]

\subsection{Robust scale estimators} \label{SEC:breakdown-scale}
For robust scale estimation,
we consider the MAD \citep{Hampel:1974} and the Shamos estimator \citep{Shamos:1976}.
The MAD is given by
\begin{equation} \label{EQ:MAD}
\mathrm{MAD} =
\frac{\displaystyle{\mathop\mathrm{median}_{1\le i\le n}}|X_i-\tilde{\mu}|}{\Phi^{-1}({3}/{4})}
\approx 1.4826\cdot\displaystyle{\mathop\mathrm{median}_{1\le i\le n}}|X_i-\tilde{\mu}|,
\end{equation}
where $\tilde{\mu} = \mathrm{median}(X_i)$ and
$\Phi^{-1}({3}/{4})$ is needed to make this estimator
Fisher-consistent \citep{Fisher:1922} for the standard deviation under the normal distribution.
For more details, see Example~4 in Section 2.3 of \cite{Hampel/etc:1986} and \cite{Rousseeuw/Croux:1992,Rousseeuw/Croux:1993}.
This resembles the median and its finite-sample breakdown point is the same as
that of the median in (\ref{EQ:e-median}). The Shamos estimator is given by
\begin{equation} \label{EQ:Shamos}
\mathrm{Shamos} =
\frac{\displaystyle\mathop{\mathrm{median}}_{i < j} \big( |X_i-X_j| \big)}%
{\sqrt{2}\,\Phi^{-1}(3/4)}
\approx
1.048358\cdot\displaystyle\mathop{\mathrm{median}}_{i < j} \big( |X_i-X_j| \big),
\end{equation}
where $\sqrt{2}\,\Phi^{-1}(3/4)$ is needed to make this estimator Fisher-consistent for the standard deviation
under the normal distribution \citep{Levy/etc:2011}.

\begin{figure}[t]
\centering
\includegraphics{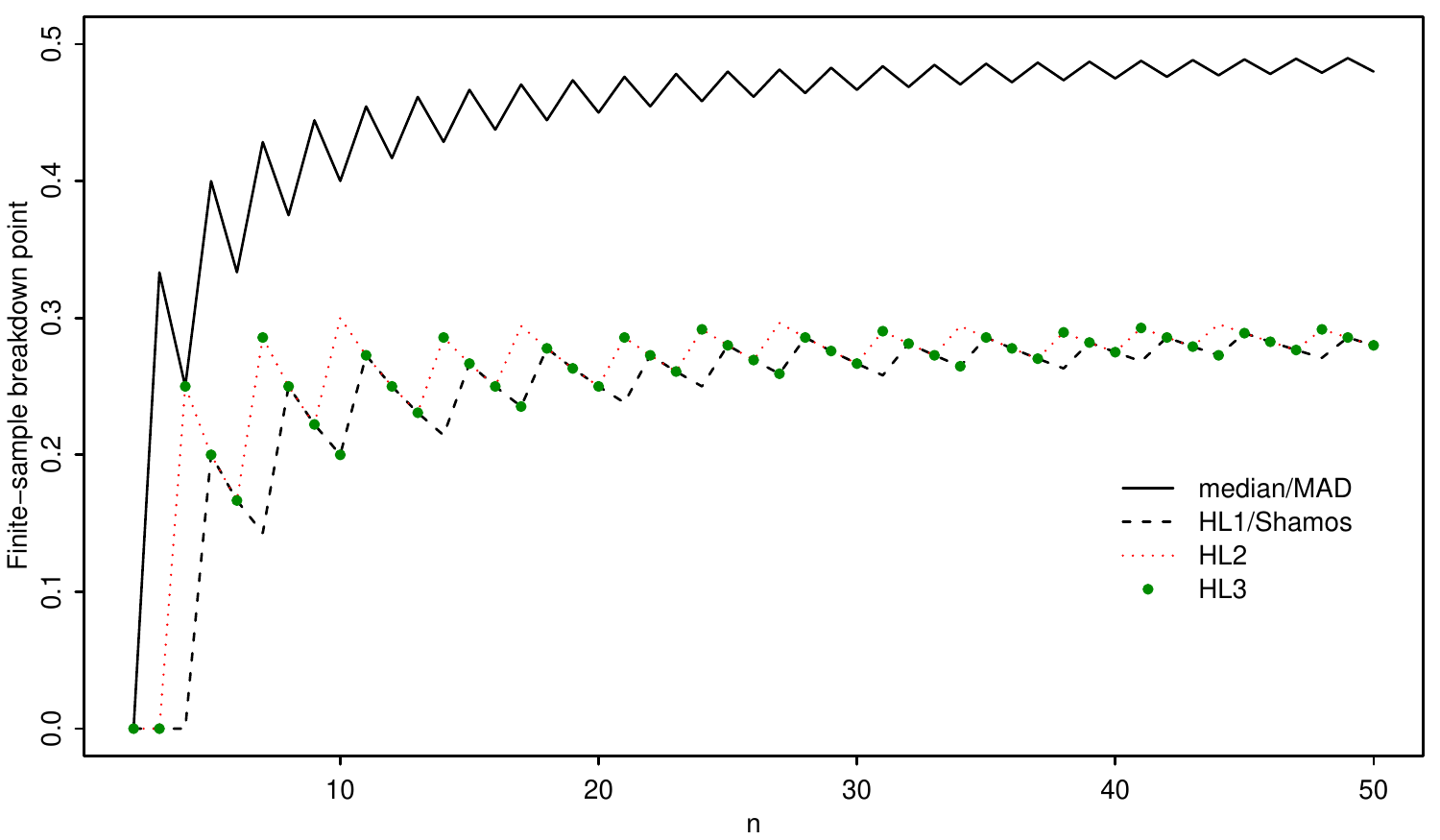}
\caption{The finite-sample breakdown points under consideration.\label{FIG:breakdown}}
\end{figure}

Of particular note is that the Shamos estimator resembles
the HL1 estimator by replacing the Walsh averages by
pairwise differences. Thus, its finite-sample breakdown point is the same
as that of the HL1 estimator in  (\ref{EQ:e-HL1}).
In the case of the HL estimator, the median is calculated
for $i<j$, $i\le j$, and $\forall(i,j)$, but the median in the Shamos estimator
is calculated only for  $i<j$ because $|X_i-X_j|=0$ for $i=j$.
Note that the MAD and Shamos estimators are in a closed-form
and are scale-equivariant in the sense that
\[
\hat{\theta}(aX_1+b,aX_2+b,\ldots,aX_n+b) = |a|\cdot\hat{\theta}(X_1,X_2,\ldots,X_n).
\]
As afore-mentioned,
we also provide the values the finite-sample breakdown points
of the Shamos estimator in Table~\ref{TBL:breakdown} and
the plot of these values in Figure~\ref{FIG:breakdown}.

\section{Empirical biases} \label{SEC:biase}
As mentioned above, the MAD in (\ref{EQ:MAD})
and the Shamos estimator in (\ref{EQ:Shamos})
are Fisher-consistent for the standard deviation under the normal
distribution, that is, as the sample size goes to infinity, it converges to
the standard deviation $\sigma$ under the normal distribution, $N(\mu,\sigma)$.
However, due to the cost of sampling and measurements in practical applications, we often need to find parameter estimates from data with small sample sizes; see, for example, \cite{Williams:2011}, whereas these estimates have serious biases, especially when the sample sizes are small.
To the best of our knowledge,
an empirical investigation of finite sample correction factors of these estimates has not been well studied in the literature. This observation motivates us to obtain the unbiasing factors for the MAD and Shamos estimators
through the Monte Carlo simulation methods \citep{Rousseeuw/Croux:1992}. In what follows, we discuss how to conduct simulations under two cases: $2 \leq n \leq 100$ and $n > 100$.
	
\begin{figure}[t]
\centering
\includegraphics{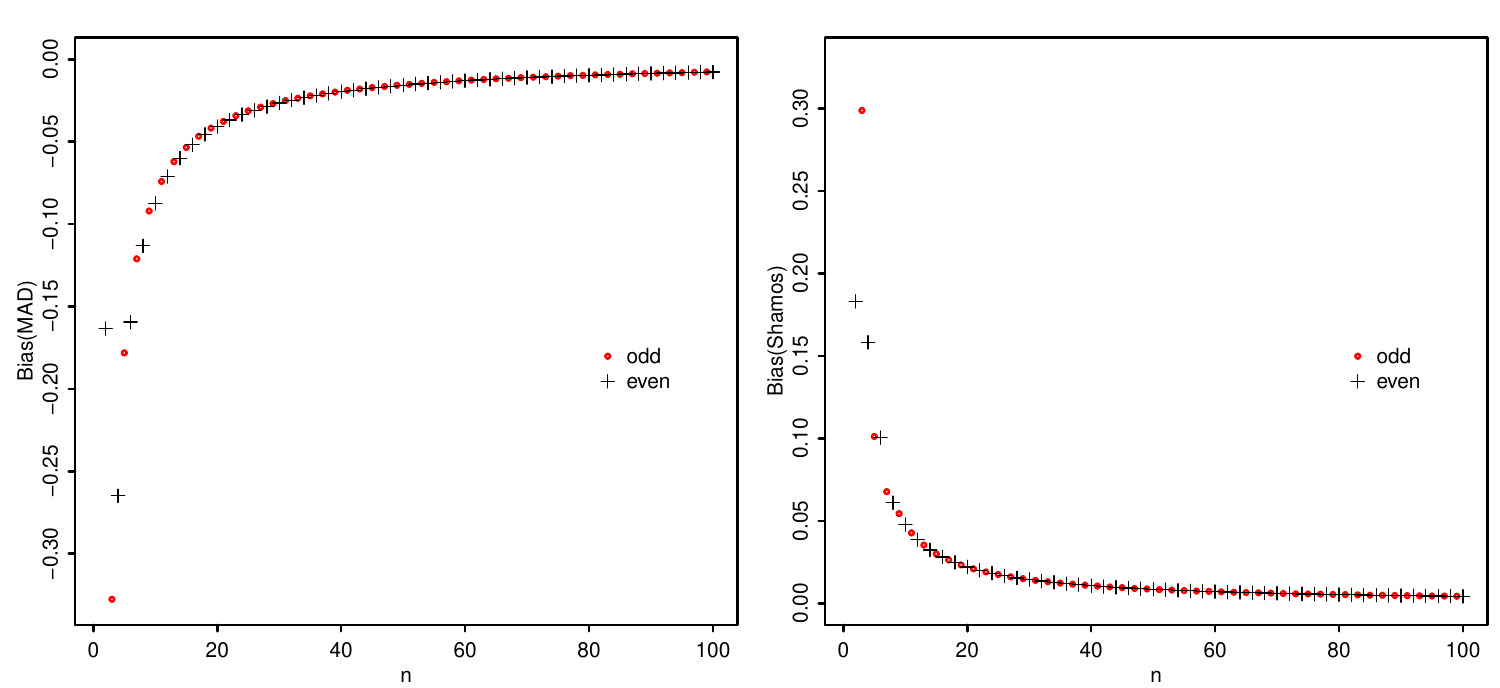}
\caption{Empirical biases of the MAD and Shamos estimators.\label{FIG:bias}}
\end{figure}

For the case when $2 \leq n \leq 100$, we generated a sample of size $n$ from the standard normal distribution,
$N(0,1)$, and calculated the MAD and Shamos estimates.
We repeated this simulation ten million times ($I=10^7$)
to obtain the empirical biases of these two estimators. These values are provided in Table~\ref{TBL:eBias1} and can also be seen in Figure~\ref{FIG:bias}. It is clear that an empirically unbiased
MAD is given by
\[
\frac{\mathrm{MAD}}{1+A_n},
\]
where $A_n$ is the empirical bias of the $\mathrm{MAD}$
under the standard normal distribution. Similarly, an empirically unbiased Shamos estimator is given by
\[
\frac{\mathrm{Shamos}}{1+B_n},
\]
where $B_n$ is the empirical bias of the Shamos estimator
under the standard normal distribution.
	
For the case when $n > 100$, we suggested to use the
methods proposed by \cite{Hayes:2014} and \cite{Williams:2011}. To be more specific, \cite{Hayes:2014} suggested the use of
$A_n \approx  {a_1}/{n} + {a_2}/{n^2}$
and \cite{Williams:2011} suggested the use of
$A_n \approx  {a_3} {n}^{-a_4}$. Similarly, we can estimate $B_n$ using
$B_n \approx  {b_1}/{n} + {b_2}/{n^2}$
and
$B_n \approx  {b_3} {n}^{-b_4}$.
These suggestions are quite reasonable,
because the MAD in (\ref{EQ:MAD}) and Shamos in (\ref{EQ:Shamos}) are Fisher-consistent,
and $A_n$ and $B_n$ converge to zero as $n$ goes to infinity.
	
Using the methods proposed by \cite{Hayes:2014} and \cite{Williams:2011},
we obtained the least squares estimate $A_n$ as follows:
\[
A_n = - \frac{0.76213}{n} - \frac{0.86413}{n^2},
\]
and
\[
A_n =  -0.804168866 \cdot n^{-1.008922},
\]
respectively. In a similar fashion,  the least squares estimate $B_n$ using \cite{Hayes:2014} and
\cite{Williams:2011} is given by
\begin{align*}
B_n &= \frac{0.414253297}{n} + \frac{0.442396799}{n^2}  \\
\intertext{and}
B_n & = 0.435760656 \cdot n^{-1.0084443},
\end{align*}
respectively.  The empirical biases of the MAD and Shamos estimators are provided in Table~\ref{TBL:eBias2} for
$n=109, 110, 119, 120, \ldots, 499, 500$.
We observed that the estimated biases are very accurate up to the fourth decimal point.
Note that \cite{Hayes:2014} and \cite{Williams:2011} estimated $A_n$ for the case of
odd and even values of $n$ separately.
However, as one expects, for a large value of $n$, the gain in precision may not be noticeable
as shown in Figure~\ref{FIG:bias}.

It is well known that the sample standard deviation $S_n$
is not unbiased under the normal distribution.
To make it unbiased, the unbiasing factor $c_4(n)$ is widely used in engineering statistics
	including control charting so that $S_n/c_4(n)$ is unbiased, where $c_4(n)$ is given by
\begin{equation} \label{EQ:c4}
c_4(n) = \sqrt{\frac{2}{n-1}} \cdot \frac{\Gamma(n/2)}{\Gamma(n/2-1/2)} .
\end{equation}
It is worth mentioning that this $c_4$ notation was originally used in \cite{ASTM:1976} 
and it has since then been a standard notation in the engineering statistics literature.

We suggest to use $c_5(n)$ and $c_6(n)$ notations
for the unbiasing factors of the MAD and Shamos estimators, respectively.
Consequently, we can obtain the unbiased MAD and Shamos estimators for any value of $n$ given by
\begin{equation} \label{EQ:unbiased-MAD-Shamos}
\frac{\mathrm{MAD}}{c_5(n)} \text{~~and~~} \frac{\mathrm{Shamos}}{c_6(n)},
\end{equation}
where $c_5(n)=1+A_n$ and $c_6(n)=1+B_n$.

\section{Empirical variances and relative efficiencies} \label{SEC:empirical}
In this section, through the extensive Monte Carlo simulations,
we calculate the finite-sample variances of the median, Hodges-Lehmann (HL1, HL2, HL3),
MAD and Shamos estimators under the standard normal distribution.
We generated a sample of size $n$ from the standard
normal distribution and calculated their empirical variances for a given value of $n$.
We repeated this simulation ten million times ($I=10^7$)
to obtain the empirical variance for each of $n=2, 3, \ldots, 100$.

It should be noted that the values of the asymptotic relative efficiency (ARE)
of various estimators are known.
Here, the ARE is defined as the limit of the relative efficiency (RE) as $n$ goes to infinity.
The RE of $\hat{\theta}_2$ with respect to $\hat{\theta}_1$ is defined as
\[
\mathrm{RE}(\hat{\theta}_2 | \hat{\theta}_1) =
\frac{\mathrm{Var}(\hat{\theta}_1)}{\mathrm{Var}(\hat{\theta}_2)} .
\]
where $\hat{\theta}_1$ is often a reference or baseline estimator.
Then the ARE is given by
\[
\mathrm{ARE}(\hat{\theta}_2 | \hat{\theta}_1)
= \lim_{n\to\infty} \mathrm{RE}(\hat{\theta}_2 | \hat{\theta}_1) .
\]
For example, under the normal distribution, we have
$\mathrm{ARE}(\mathrm{median} | \bar{X}) = {2}/{\pi} \approx 0.6366 $,
$\mathrm{ARE}(\mathrm{HL} | \bar{X}) = {3}/{\pi} \approx  0.9549$,
$\mathrm{ARE}(\mathrm{MAD} | S_n) = 0.37$, and
$\mathrm{ARE}(\mathrm{Shamos} | S_n) = 0.863$,
where $\bar{X}$ is the sample mean and $S_n$ is the sample standard deviation.
For more details, see \cite{Serfling:2011} and \cite{Levy/etc:2011}.

Note that with a random sample of size $n$ from the standard normal distribution,
we have $\mathrm{Var}(\bar{X})=1/n$ and
$\mathrm{Var}(S_n)=1- c_4(n)^2$,
where $c_4(n)$ is given in (\ref{EQ:c4}).
Thus, we have
\begin{align*}
n\,\mathrm{Var}(\mathrm{median}) &\approx \frac{1}{0.6366} = 1.5700 \\
n\,\mathrm{Var}(\mathrm{HL})     &\approx \frac{1}{0.9549} = 1.0472 \\
\frac{\mathrm{Var}(\mathrm{MAD})}{1-c_4(n)^2}  &\approx \frac{1}{0.3700} = 2.7027 \\
\intertext{and}
\frac{\mathrm{Var}(\mathrm{Shamos})}{1-c_4(n)^2} &\approx \frac{1}{0.8630} = 1.15875
\end{align*}
for a large value of $n$. We provided these values for each of $n$
in Tables~\ref{TBL:nvar1} and \ref{TBL:nvar2} and also plotted these values in Figure~\ref{FIG:nvar}.

For the case when  $n > 100$, we suggest to estimate these values based on \cite{Hayes:2014} or \cite{Williams:2011} as we did the biases in the previous section. We suggest the following models to obtain these values for $n>100$:
\begin{align*}
n\,\mathrm{Var}(\mathrm{median}) &= 1.5700 + \frac{a_1}{n} + \frac{a_2}{n^2}  \\
n\,\mathrm{Var}(\mathrm{HL})     &= 1.0472 + \frac{a_3}{n} + \frac{a_4}{n^2}  \\
\frac{\mathrm{Var}(\mathrm{MAD})}{1-c_4(n)^2}   &= 2.7027    + \frac{a_5}{n} + \frac{a_6}{n^2}  \\
\intertext{and}
\frac{\mathrm{Var}(\mathrm{Shamos})}{1-c_4(n)^2}&= 1.15875+\frac{a_7}{n}+\frac{a_8}{n^2}.
\end{align*}
One can also use the method based on \cite{Williams:2011}.
For brevity, we used the method based on \cite{Hayes:2014}.
To estimate these values for $n>100$,
we obtained the empirical REs in Table~\ref{TBL:nvar3} for
$n=109$, $110$, $119$, $120$, $\ldots$, $499$, $500$.
We observe from Figure~\ref{FIG:nvar} that it is reasonable to estimate
the values for the median and MAD in the case of odd and even values of $n$ separately.
Using the large values of $n$, we can estimate the above coefficients.
For this reason, we use the simulation results in Tables \ref{TBL:nvar2} and \ref{TBL:nvar3}.
Then the least squares estimates based on the method of \cite{Hayes:2014} are given by
\begin{align*}
n\,\mathrm{Var}(\mathrm{median})  &= 1.5700 - \frac{0.6589}{n} - \frac{0.943}{n^2}
\qquad (\textrm{for~odd~} n) \\
n\,\mathrm{Var}(\mathrm{median})  &= 1.5700 - \frac{2.1950}{n} + \frac{1.929}{n^2}
\qquad (\textrm{for~even~} n) \\
n\,\mathrm{Var}(\mathrm{HL1})     &= 1.0472 + \frac{0.1127}{n} + \frac{0.8365}{n^2}  \\
n\,\mathrm{Var}(\mathrm{HL2})     &= 1.0472 + \frac{0.2923}{n} + \frac{0.2258}{n^2}  \\
n\,\mathrm{Var}(\mathrm{HL3})     &= 1.0472 + \frac{0.2022}{n} + \frac{0.4343}{n^2}  \\
\frac{\mathrm{Var}(\mathrm{MAD})}{1-c_4(n)^2} &= 2.7027 +\frac{0.2996}{n} - \frac{149.357}{n^2}
\qquad (\textrm{for~odd~} n) \\
\frac{\mathrm{Var}(\mathrm{MAD})}{1-c_4(n)^2} &= 2.7027 - \frac{2.417}{n} - \frac{153.010}{n^2}
\qquad (\textrm{for~even~} n)
\intertext{and}
\frac{\mathrm{Var}(\mathrm{Shamos})}{1-c_4(n)^2}&
= 1.15875 +\frac{2.822}{n} +\frac{12.238}{n^2}.
\end{align*}

In Tables~\ref{TBL:RE1} and \ref{TBL:RE2}, we also calculated the REs of
the afore-mentioned estimators for $n=1,2,\ldots,100$ using the above
empirical variances.
For $n>100$, we can easily obtain the REs using the above estimated variances.
It should be noted that the REs of the median, HL2 and HL3
are exactly one for $n=1,2$.
When $n=1,2$, the median, HL2 and HL3 are essentially
the same as the sample mean.

Another noticeable result is that the RE of the HL1 is exactly one especially for $n=4$.
When $n=4$, the HL1 is the median of
$(X_1+X_2)/2$, $(X_1+X_3)/2$, $(X_1+X_4)/2$,
$(X_2+X_3)/2$, $(X_2+X_4)/2$ and $(X_3+X_4)/2$.
Then this is the same as the median of
$(X_{(1)}+X_{(2)})/2$, $(X_{(1)}+X_{(3)})/2$, $(X_{(1)}+X_{(4)})/2$,
$(X_{(2)}+X_{(3)})/2$, $(X_{(2)}+X_{(4)})/2$ and $(X_{(3)}+X_{(4)})/2$,
where $X_{(i)}$ are order statistics.
Because
\begin{align*}
\frac{X_{(1)}+X_{(2)}}{2} \le \frac{X_{(1)}+X_{(3)}}{2} \le \frac{X_{(1)}+X_{(4)}}{2}
\intertext{and}
\frac{X_{(2)}+X_{(3)}}{2} \le \frac{X_{(2)}+X_{(4)}}{2} \le \frac{X_{(3)}+X_{(4)}}{2},
\end{align*}
we have
\[
\mathrm{HL}_1 =
\frac{1}{2}\left( \frac{X_{(1)}+X_{(4)}}{2}+\frac{X_{(2)}+X_{(3)}}{2} \right)
= \frac{X_1+X_2+X_3+X_4}{4} = \bar{X}.
\]
Thus, the RE of the HL1 should be exactly one.
In this case, as expected, the finite-sample breakdown
is zero as provided in Table~\ref{TBL:breakdown}.

It should be noted that the $\mathrm{MAD}/c_5(n)$ and  $\mathrm{Shamos}/c_6(n)$ are
unbiased for $\sigma$ under the normal distribution, but their square values
are not unbiased for  $\sigma^2$.
Using the empirical and estimated variances, we can obtain the unbiased
versions as follows.
For convenience, we denote
$v_5(n) = \mathrm{Var}(\mathrm{MAD})$ and
$v_6(n) = \mathrm{Var}(\mathrm{Shamos})$,
where the variances are obtained using a sample of size $n$ from the standard
normal distribution $N(0,1)$ as mentioned earlier.
Since the MAD and Shamos estimators are scale-equivariant, we have
$\mathrm{Var}(\mathrm{MAD}) = v_5(n) \sigma^2$ and
$\mathrm{Var}(\mathrm{Shamos}) = v_6(n) \sigma^2$ with a sample
from the normal distribution $N(\mu, \sigma^2)$.
It is immediate from Equation (\ref{EQ:unbiased-MAD-Shamos}) that
$E(\mathrm{MAD}) = c_5(n) \sigma$ and
$E(\mathrm{Shamos}) = c_6(n) \sigma$.
Considering $E(\hat{\theta}^2) = \mathrm{Var}(\hat{\theta}) + E(\hat{\theta})^2$,
we have
\begin{align*}
E(\mathrm{MAD}^2)    &= v_5(n) \sigma^2 +  \big\{ c_5(n) \sigma \big\}^2 \\
\intertext{and}
E(\mathrm{Shamos}^2) &= v_6(n) \sigma^2 +  \big\{ c_6(n) \sigma \big\}^2 .
\end{align*}
Thus, the following estimators are unbiased for $\sigma^2$ under the normal distribution
\[
\frac{\mathrm{MAD}^2}{v_5(n) + c_5(n)^2} \text{~~~and~~~}
\frac{\mathrm{Shamos}^2}{v_6(n) + c_6(n)^2} .
\]

\section{{Example}} \label{section:Example}

In this section, we provide an example to illustrate the
application of the proposed methods to statistical quality control charts \citep{Shewhart:1926b}
for averages and standard deviations.
For more details, see Section 3.8 of \cite{ASTM:2018}.
Three-sigma control limits are most widely used
to set the upper and lower control limits in statistical control charts due to \cite{Shewhart:1931}.

We briefly introduce how to construct the statistical control charts
and then propose robust alternatives using the proposed methods.
In general, the construction of statistical quality control charts is involved with
two phases: Phase-I and Phase-II \citep{Vining:2009,Montgomery:2013a}.
In Phase-I, the goal is to establish reliable control limits with a clean set of
process data. In Phase-II, using the control limits obtained in Phase-I, we monitor
the process by comparing the statistic for each successive subgroup as future observations
are collected.
The Phase-II control charts are based on the estimates in Phase-I and
the performance of the Phase-II control chart relies on the accuracy of the Phase-I estimate.
Thus, if the Phase-I control limits are estimated with a corrupted data set,
this can produce an adverse effect in Phase-II.
It is well known that the sample mean and standard deviations are highly sensitive
to data contamination and can even be destroyed by a single corrupted value \citep{Rousseeuw:1991}.
Thus, the conventional control charts based on these statistics
are very sensitive to contamination.
This problem can be easily overcome by using the proposed robust methods.

As an illustration, we deal with the control limits of the conventional control charts
for averages and standard deviations.
Suppose that the mean and standard deviation are given by $\mu$ and $\sigma$ with
a sample of size $n$.
Then the parameters of the control limits of the $\bar{x}$ chart are given by
$\mu+3\sigma/\sqrt{n}$ (UCL) and $\mu-3\sigma/\sqrt{n}$ (LCL).
We assume that we have $k$ subgroups which is of size $n$.
Let $X_{ij}$ be the $i$th sample, where $i=1,2,\ldots,k$ and $j=1,2,\ldots,n$
and $x_{ij}$ be its realization.
Let $\bar{x}_i$ and $s_i$ be the sample mean and standard deviation of the $i$th sample,
respectively.
Then the estimates of $\mu \pm 3\sigma/\sqrt{n}$ are given by
$\bar{\bar{x}} \pm A_3(n) \bar{s}$,
where $\bar{\bar{x}} = \sum_{i=1}^k \bar{x}_i/k$,
$\bar{s} = \sum_{i=1}^k s_i/k$, and $A_3(n) = 3/c_4(n)\sqrt{n}$;
see \cite{ASQC-A1:1972} for the notation $A_3(n)$.
Note that an estimate of $3\sigma$ is given by $\sqrt{n} A_3(n) \bar{s}$ since $A_3(n) \bar{s}$ is an estimate of $3\sigma/\sqrt{n}$.

As obtained in (\ref{EQ:unbiased-MAD-Shamos}),
the unbiased MAD and Shamos estimators for any value of $n$ are given by
\[
\frac{\mathrm{MAD}}{c_5(n)} \text{~~and~~} \frac{\mathrm{Shamos}}{c_6(n)},
\]
respectively. For convenience, we let  $A_5(n) = 3/c_5(n)\sqrt{n}$ and $A_6(n) = 3/c_6(n)\sqrt{n}$.
Let $\mathrm{MAD}_i$ and $\mathrm{SH}_i$ be
the MAD and Shamos estimates of the $i$th sample, respectively.
Then robust analogues of $3\sigma$ can be obtained as
\[
\sqrt{n} \cdot A_5(n) \cdot \overline{\mathrm{MAD}} \textrm{~~and~~}
\sqrt{n} \cdot A_6(n) \cdot \overline{\mathrm{SH}},
\]
where
$\overline{\mathrm{MAD}} = \sum_{i=1}^k \mathrm{MAD}_i/k$ and
$\overline{\mathrm{SH}} = \sum_{i=1}^k \mathrm{SH}_i/k$.

As mentioned above, the conventional limits are very sensitive to data contamination.
For example, when there are outlying observations, the control limits
are moved outward from the center line, which makes it difficult to monitor
any out-of-control signal.
Since the \emph{three-sigma} plays an important role in constructing the charts,
we investigate the estimates of  $3\sigma$.
To this end, we carry out a simulation.
We assume that $X_{ij}$ are normally distributed with $N(\mu,\sigma^2)$ where $\mu=5$ and $\sigma=1$
and there are $k=10$ subgroups which are all of size $n=5$.
We repeated this simulation ten thousand times.

\begin{figure}[tbh]
	\centering
	\includegraphics{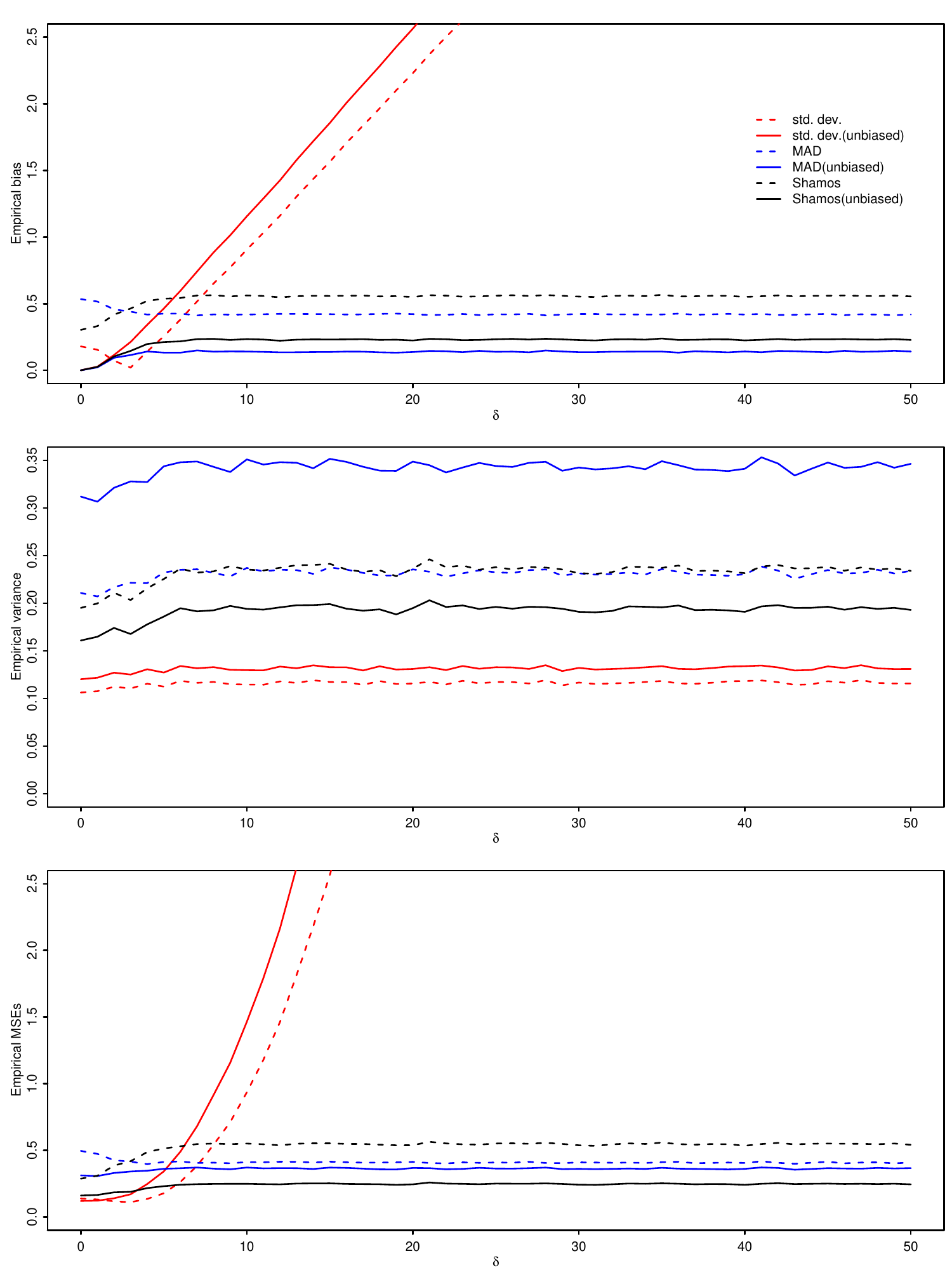}
	\caption{The empirical biases, variances and mean square errors of
		the estimates under consideration.\label{FIG:EXplot}}
\end{figure}

To investigate the robust properties of the proposed methods,
we corrupted the observation $X_{11}$ by adding $\delta$ to $X_{11}$, where $\delta=0, 1, 2, \ldots, 50$.
We provide the empirical biases, variances and mean square errors of the estimates of $3\sigma$
in Table~\ref{TBL:BiasVarMSE} and Figure~\ref{FIG:EXplot}.
For brevity, we provide the results only for $\delta=0,10,20,30,40,50$ in the table.
In the figure, the solid lines denote the unbiased estimates after the finite-sample correction.
From the results, if there is no contamination ($\delta=0$), the unbiased/original standard deviations
clearly perform the best. The unbiased version is slightly better than the original standard deviation.
Thus, the unbiasing factor, $c_4(n)$, improves the bias and MSE.
However, if there exists data contamination, the unbiased/original standard deviations started to break down
as the corruption level $\delta$ increases.
We also observe the unbiasing factor, $c_4(n)$, does not improve either the bias or the MSE
unless $\delta$ is very small.

On the other hand, the robust alternatives (MAD and Shamos) do not break down with the considered corruption as expected.
We also note that the unbiasing factors, $c_5(n)$ and $c_6(n)$, could help the bias and MSE decrease
even when there is contamination.
Especially, the original Shamos estimator is much improved with the unbiasing factor $c_6(n)$.
Note that the MSE of the unbiased standard deviation is $0.12023$, while that of the unbiased Shamos
is $0.16093$ when there is no contamination. This shows that the two estimators are quite comparable
even when there is no contamination.
In summary, the unbiased Shamos is the best choice in this illustrative example.


\begin{table}[htb]
	\caption{Empirical biases, variances and mean square errors of the estimates under consideration.
		\label{TBL:BiasVarMSE}}
	\medskip
	\begin{center}
		\begin{tabular}{clrrrrrr} 
			\hline
			&& standard  & unbiased   &         & unbiased &        & unbiased \\
			$\delta$ && deviation & std.\ dev. &   MAD   & MAD      & Shamos & Shamos  \\
			\cline{1-1} \cline{3-8}
			&& \multicolumn{6}{c}{\emph{Empirical Biases}}   \\
			0&& $-0.17971$&   0.00036 &  $-0.53401$ &   0.00044 &   0.30333 & $-0.00018$ \\
			10&&   0.90590 &   1.15528 &  $-0.41840$ &   0.14111 &   0.56179 &   0.23454  \\
			20&&   2.22951 &   2.56340 &  $-0.42168$ &   0.13713 &   0.54997 &   0.22381  \\
			30&&   3.57069 &   3.99020 &  $-0.42252$ &   0.13610 &   0.55349 &   0.22700  \\
			40&&   4.91276 &   5.41796 &  $-0.41812$ &   0.14145 &   0.55059 &   0.22436  \\
			50&&   6.25617 &   6.84714 &  $-0.41834$ &   0.14119 &   0.55455 &   0.22796  \\
			\cline{1-1} \cline{3-8}
			&& \multicolumn{6}{c}{\emph{Empirical Variances}}   \\
			0&&  0.10623 &  0.12023 &  0.21083 &  0.31212 &  0.19514 &  0.16093   \\
			10&&  0.11460 &  0.12970 &  0.23713 &  0.35106 &  0.23538 &  0.19411   \\
			20&&  0.11573 &  0.13098 &  0.23563 &  0.34884 &  0.23632 &  0.19489   \\
			30&&  0.11673 &  0.13211 &  0.23140 &  0.34258 &  0.23159 &  0.19099   \\
			40&&  0.11833 &  0.13392 &  0.23053 &  0.34129 &  0.23157 &  0.19098   \\
			50&&  0.11572 &  0.13097 &  0.23405 &  0.34649 &  0.23397 &  0.19295   \\
			\cline{1-1} \cline{3-8}
			&& \multicolumn{6}{c}{\emph{Empirical MSEs}}   \\
			0&&   0.13853 &  0.12023 &  0.49600 &  0.31212 &  0.28715 &  0.16093 \\
			10&&   0.93526 &  1.46437 &  0.41219 &  0.37097 &  0.55099 &  0.24912 \\
			20&&   5.08646 &  6.70198 &  0.41344 &  0.36764 &  0.53879 &  0.24498 \\
			30&&  12.86655 & 16.05382 &  0.40993 &  0.36110 &  0.53794 &  0.24252 \\
			40&&  24.25359 & 29.48825 &  0.40536 &  0.36130 &  0.53472 &  0.24131 \\
			50&&  39.25542 & 47.01434 &  0.40905 &  0.36643 &  0.54149 &  0.24492 \\
			\hline
		\end{tabular}
	\end{center}

\end{table}
	
\clearpage
\section{{Concluding remarks}} \label{section:Concluding}
In this paper, we studied the finite-sample properties of
the sample median and Hodges-Lehmann estimators for location
and the MAD and Shamos estimators for scale.
We obtained the closed-form finite-sample breakdown points of these estimators for the population parameters.
We also calculated the unbiasing factors and relative efficiencies of
the MAD and the Shamos estimators for the scale parameter
through the extensive Monte Carlo simulations up to the sample size of 100.
The numerical study showed that the unbiasing factor
significantly improves the finite-sample performance of these estimators under consideration.
In addition, we provided the predicted values for the
unbiasing factors obtained by using the least squares method
which can be used for the case of sample size more than 100.
To facilitate the implementation of the proposed methods into various application fields, \cite{Park/Wang:2019b}  developed the R package library
which is published in the CRAN \citep{R:2019}.
	
According to the application of the proposed methods to statistical quality control charts for averages and standard deviations,
it would be useful to develop robustified alternatives to
traditional Shewhart-type control charts
because its control limits are also constructed based on the unbiased estimators under the normal distribution.
We obtained the empirical and estimated biases of several commonly used robust estimators under the normal distribution,
which can be widely used for bias-corrections  of the estimates of the parameters. For other parametric models, one can obtain
empirical and estimated biases with appropriate robust estimators as suggested in this paper.
Given that the breakdown point used in this paper
does not depend on a probability model as pointed out in Subsection 2.2a
of \cite{Hampel/etc:1986}, it is of particular interest to obtain biases and relative efficiencies under other parametric models.

\section*{Acknowledgment}
The authors greatly appreciate the reviewers for their thoughtful comments and suggestions, which have significantly improved the quality of this manuscript. This work was supported by the National Research Foundation of Korea (NRF) grant funded by the Korea government (No. NRF-2017R1A2B4004169).

\bibliographystyle{apalike}
\bibliography{Reference}

\clearpage
\appendix
\vspace*{2in}
\section*{\huge Appendix: Tables and Figures}

\begin{table}[htb]
	\caption{Finite-sample breakdown points.\label{TBL:breakdown}}
	\medskip
	\begin{spacing}{1.0}\centering
		\begin{small}
			\begin{tabular}{clcccc} 
				\hline
				$n$ && median and  MAD & HL1 and Shamos & HL2 & HL3  \\
				\cline{1-1} \cline{3-6}
				2 && 0.0000000 & 0.0000000 & 0.0000000 & 0.0000000 \\
				3 && 0.3333333 & 0.0000000 & 0.0000000 & 0.0000000 \\
				4 && 0.2500000 & 0.0000000 & 0.2500000 & 0.2500000 \\
				5 && 0.4000000 & 0.2000000 & 0.2000000 & 0.2000000 \\
				6 && 0.3333333 & 0.1666667 & 0.1666667 & 0.1666667 \\
				7 && 0.4285714 & 0.1428571 & 0.2857143 & 0.2857143 \\
				8 && 0.3750000 & 0.2500000 & 0.2500000 & 0.2500000 \\
				9 && 0.4444444 & 0.2222222 & 0.2222222 & 0.2222222 \\
				10 && 0.4000000 & 0.2000000 & 0.3000000 & 0.2000000 \\
				11 && 0.4545455 & 0.2727273 & 0.2727273 & 0.2727273 \\
				12 && 0.4166667 & 0.2500000 & 0.2500000 & 0.2500000 \\
				13 && 0.4615385 & 0.2307692 & 0.2307692 & 0.2307692 \\
				14 && 0.4285714 & 0.2142857 & 0.2857143 & 0.2857143 \\
				15 && 0.4666667 & 0.2666667 & 0.2666667 & 0.2666667 \\
				16 && 0.4375000 & 0.2500000 & 0.2500000 & 0.2500000 \\
				17 && 0.4705882 & 0.2352941 & 0.2941176 & 0.2352941 \\
				18 && 0.4444444 & 0.2777778 & 0.2777778 & 0.2777778 \\
				19 && 0.4736842 & 0.2631579 & 0.2631579 & 0.2631579 \\
				20 && 0.4500000 & 0.2500000 & 0.2500000 & 0.2500000 \\
				21 && 0.4761905 & 0.2380952 & 0.2857143 & 0.2857143 \\
				22 && 0.4545455 & 0.2727273 & 0.2727273 & 0.2727273 \\
				23 && 0.4782609 & 0.2608696 & 0.2608696 & 0.2608696 \\
				24 && 0.4583333 & 0.2500000 & 0.2916667 & 0.2916667 \\
				25 && 0.4800000 & 0.2800000 & 0.2800000 & 0.2800000 \\
				26 && 0.4615385 & 0.2692308 & 0.2692308 & 0.2692308 \\
				27 && 0.4814815 & 0.2592593 & 0.2962963 & 0.2592593 \\
				28 && 0.4642857 & 0.2857143 & 0.2857143 & 0.2857143 \\
				29 && 0.4827586 & 0.2758621 & 0.2758621 & 0.2758621 \\
				30 && 0.4666667 & 0.2666667 & 0.2666667 & 0.2666667 \\
				31 && 0.4838710 & 0.2580645 & 0.2903226 & 0.2903226 \\
				32 && 0.4687500 & 0.2812500 & 0.2812500 & 0.2812500 \\
				33 && 0.4848485 & 0.2727273 & 0.2727273 & 0.2727273 \\
				34 && 0.4705882 & 0.2647059 & 0.2941176 & 0.2647059 \\
				35 && 0.4857143 & 0.2857143 & 0.2857143 & 0.2857143 \\
				36 && 0.4722222 & 0.2777778 & 0.2777778 & 0.2777778 \\
				37 && 0.4864865 & 0.2702703 & 0.2702703 & 0.2702703 \\
				38 && 0.4736842 & 0.2631579 & 0.2894737 & 0.2894737 \\
				39 && 0.4871795 & 0.2820513 & 0.2820513 & 0.2820513 \\
				40 && 0.4750000 & 0.2750000 & 0.2750000 & 0.2750000 \\
				41 && 0.4878049 & 0.2682927 & 0.2926829 & 0.2926829 \\
				42 && 0.4761905 & 0.2857143 & 0.2857143 & 0.2857143 \\
				43 && 0.4883721 & 0.2790698 & 0.2790698 & 0.2790698 \\
				44 && 0.4772727 & 0.2727273 & 0.2954545 & 0.2727273 \\
				45 && 0.4888889 & 0.2888889 & 0.2888889 & 0.2888889 \\
				46 && 0.4782609 & 0.2826087 & 0.2826087 & 0.2826087 \\
				47 && 0.4893617 & 0.2765957 & 0.2765957 & 0.2765957 \\
				48 && 0.4791667 & 0.2708333 & 0.2916667 & 0.2916667 \\
				49 && 0.4897959 & 0.2857143 & 0.2857143 & 0.2857143 \\
				50 && 0.4800000 & 0.2800000 & 0.2800000 & 0.2800000 \\
				\hline
			\end{tabular}
		\end{small}
	\end{spacing}
\end{table}

\begin{table}[htb]
	\caption{Empirical biases of the MAD and Shamos estimators
		($n=2,3,\ldots,100$).\label{TBL:eBias1}}
	\medskip
	\begin{spacing}{1.0}\centering
		\begin{small}
			\begin{tabular}{clcclclcc} 
				\hline
				$n$ && MAD & Shamos && $n$ && MAD & Shamos  \\
				\cline{1-1} \cline{3-4} \cline{6-6} \cline{8-9}
				1 &&   NA         &   NA         &&  51 && $-0.0152820$ & $0.0082120$ \\
				2 && $-0.1633880$ & $0.1831500$  &&  52 && $-0.0149951$ & $0.0081874$ \\
				3 && $-0.3275897$ & $0.2989400$  &&  53 && $-0.0146042$ & $0.0079775$ \\
				4 && $-0.2648275$ & $0.1582782$  &&  54 && $-0.0145007$ & $0.0078126$ \\
				5 && $-0.1781250$ & $0.1011748$  &&  55 && $-0.0140391$ & $0.0076743$ \\
				6 && $-0.1594213$ & $0.1005038$  &&  56 && $-0.0139674$ & $0.0075212$ \\
				7 && $-0.1210631$ & $0.0676993$  &&  57 && $-0.0136336$ & $0.0074051$ \\
				8 && $-0.1131928$ & $0.0609574$  &&  58 && $-0.0134819$ & $0.0072528$ \\
				9 && $-0.0920658$ & $0.0543760$  &&  59 && $-0.0130812$ & $0.0071807$ \\
				10 && $-0.0874503$ & $0.0476839$  &&  60 && $-0.0129708$ & $0.0070617$ \\
				11 && $-0.0741303$ & $0.0426722$  &&  61 && $-0.0126589$ & $0.0069123$ \\
				12 && $-0.0711412$ & $0.0385003$  &&  62 && $-0.0125598$ & $0.0067833$ \\
				13 && $-0.0620918$ & $0.0353028$  &&  63 && $-0.0122696$ & $0.0066439$ \\
				14 && $-0.0600210$ & $0.0323526$  &&  64 && $-0.0121523$ & $0.0065821$ \\
				15 && $-0.0534603$ & $0.0299677$  &&  65 && $-0.0118163$ & $0.0064889$ \\
				16 && $-0.0519047$ & $0.0280421$  &&  66 && $-0.0118244$ & $0.0063844$ \\
				17 && $-0.0467319$ & $0.0262195$  &&  67 && $-0.0115177$ & $0.0062930$ \\
				18 && $-0.0455579$ & $0.0247674$  &&  68 && $-0.0114479$ & $0.0061910$ \\
				19 && $-0.0417554$ & $0.0232297$  &&  69 && $-0.0111309$ & $0.0061255$ \\
				20 && $-0.0408248$ & $0.0220155$  &&  70 && $-0.0110816$ & $0.0060681$ \\
				21 && $-0.0376967$ & $0.0208687$  &&  71 && $-0.0108875$ & $0.0058994$ \\
				22 && $-0.0368350$ & $0.0199446$  &&  72 && $-0.0108319$ & $0.0058235$ \\
				23 && $-0.0342394$ & $0.0189794$  &&  73 && $-0.0106032$ & $0.0057172$ \\
				24 && $-0.0335390$ & $0.0182343$  &&  74 && $-0.0105424$ & $0.0056805$ \\
				25 && $-0.0313065$ & $0.0174421$  &&  75 && $-0.0102237$ & $0.0056343$ \\
				26 && $-0.0309765$ & $0.0166364$  &&  76 && $-0.0102132$ & $0.0055605$ \\
				27 && $-0.0290220$ & $0.0160158$  &&  77 && $-0.0099408$ & $0.0055011$ \\
				28 && $-0.0287074$ & $0.0153715$  &&  78 && $-0.0099776$ & $0.0053872$ \\
				29 && $-0.0269133$ & $0.0148940$  &&  79 && $-0.0097815$ & $0.0053062$ \\
				30 && $-0.0265451$ & $0.0144027$  &&  80 && $-0.0097399$ & $0.0052348$ \\
				31 && $-0.0250734$ & $0.0138855$  &&  81 && $-0.0094837$ & $0.0052075$ \\
				32 && $-0.0248177$ & $0.0134510$  &&  82 && $-0.0094713$ & $0.0051173$ \\
				33 && $-0.0236460$ & $0.0130228$  &&  83 && $-0.0092390$ & $0.0050697$ \\
				34 && $-0.0232808$ & $0.0127183$  &&  84 && $-0.0092875$ & $0.0049805$ \\
				35 && $-0.0222099$ & $0.0122444$  &&  85 && $-0.0091508$ & $0.0048705$ \\
				36 && $-0.0220756$ & $0.0118214$  &&  86 && $-0.0090145$ & $0.0048695$ \\
				37 && $-0.0210129$ & $0.0115469$  &&  87 && $-0.0088191$ & $0.0048287$ \\
				38 && $-0.0207309$ & $0.0113206$  &&  88 && $-0.0088205$ & $0.0047315$ \\
				39 && $-0.0199272$ & $0.0109636$  &&  89 && $-0.0086622$ & $0.0046961$ \\
				40 && $-0.0197140$ & $0.0106308$  &&  90 && $-0.0085714$ & $0.0046698$ \\
				41 && $-0.0188446$ & $0.0104384$  &&  91 && $-0.0084718$ & $0.0046010$ \\
				42 && $-0.0188203$ & $0.0100693$  &&  92 && $-0.0083861$ & $0.0045544$ \\
				43 && $-0.0180521$ & $0.0098523$  &&  93 && $-0.0082559$ & $0.0045191$ \\
				44 && $-0.0178185$ & $0.0096735$  &&  94 && $-0.0082650$ & $0.0044245$ \\
				45 && $-0.0171866$ & $0.0094973$  &&  95 && $-0.0080977$ & $0.0044074$ \\
				46 && $-0.0170796$ & $0.0092210$  &&  96 && $-0.0080708$ & $0.0043579$ \\
				47 && $-0.0165391$ & $0.0089781$  &&  97 && $-0.0078810$ & $0.0043536$ \\
				48 && $-0.0163509$ & $0.0088083$  &&  98 && $-0.0078492$ & $0.0042874$ \\
				49 && $-0.0157862$ & $0.0086574$  &&  99 && $-0.0077043$ & $0.0042520$ \\
				50 && $-0.0157372$ & $0.0084772$  && 100 && $-0.0077614$ & $0.0041864$ \\
				\hline
			\end{tabular}
		\end{small}
	\end{spacing}
\end{table}

\begin{table}[htb]
	\caption{Empirical biases of the MAD and Shamos estimators along with their estimates
		($n=109,110, 119, 120, \ldots,499, 500$).\label{TBL:eBias2}}
	\medskip
	\begin{spacing}{1.0}\centering
		\begin{small}
			\begin{tabular}{clccclccc} 
				\hline
				$n$ && MAD & $A_n$(Hayes) & $A_n$(Williams) && Shamos & $B_n$(Hayes) & $B_n$(Williams)  \\
				\cline{1-1} \cline{3-5} \cline{7-9}
				109 && $-0.0070577$ & $-0.0070648$ & $-0.0070753$ && 0.0038374 & 0.0038377 & 0.0038425  \\
				110 && $-0.0070262$ & $-0.0069999$ & $-0.0070104$ && 0.0037996 & 0.0038025 & 0.0038073  \\
				119 && $-0.0064893$ & $-0.0064655$ & $-0.0064756$ && 0.0034984 & 0.0035124 & 0.0035170  \\
				120 && $-0.0064342$ & $-0.0064111$ & $-0.0064212$ && 0.0034691 & 0.0034828 & 0.0034875  \\
				129 && $-0.0059226$ & $-0.0059599$ & $-0.0059693$ && 0.0032441 & 0.0032379 & 0.0032422  \\
				130 && $-0.0059018$ & $-0.0059137$ & $-0.0059230$ && 0.0032241 & 0.0032127 & 0.0032170  \\
				139 && $-0.0054913$ & $-0.0055277$ & $-0.0055362$ && 0.0029854 & 0.0030031 & 0.0030070  \\
				140 && $-0.0055067$ & $-0.0054879$ & $-0.0054963$ && 0.0029548 & 0.0029815 & 0.0029854  \\
				149 && $-0.0051140$ & $-0.0051539$ & $-0.0051615$ && 0.0028230 & 0.0028002 & 0.0028036  \\
				150 && $-0.0051018$ & $-0.0051193$ & $-0.0051267$ && 0.0028080 & 0.0027814 & 0.0027847  \\
				159 && $-0.0047958$ & $-0.0048275$ & $-0.0048340$ && 0.0026355 & 0.0026229 & 0.0026258  \\
				160 && $-0.0047790$ & $-0.0047971$ & $-0.0048035$ && 0.0026154 & 0.0026064 & 0.0026093  \\
				169 && $-0.0045272$ & $-0.0045399$ & $-0.0045455$ && 0.0024503 & 0.0024667 & 0.0024692  \\
				170 && $-0.0045260$ & $-0.0045130$ & $-0.0045185$ && 0.0024402 & 0.0024521 & 0.0024545  \\
				179 && $-0.0042827$ & $-0.0042847$ & $-0.0042894$ && 0.0023257 & 0.0023281 & 0.0023301  \\
				180 && $-0.0042517$ & $-0.0042607$ & $-0.0042653$ && 0.0023122 & 0.0023151 & 0.0023170  \\
				189 && $-0.0040825$ & $-0.0040566$ & $-0.0040605$ && 0.0021780 & 0.0022042 & 0.0022058  \\
				190 && $-0.0040837$ & $-0.0040351$ & $-0.0040389$ && 0.0021673 & 0.0021925 & 0.0021941  \\
				199 && $-0.0038386$ & $-0.0038516$ & $-0.0038546$ && 0.0020904 & 0.0020928 & 0.0020940  \\
				200 && $-0.0038277$ & $-0.0038323$ & $-0.0038352$ && 0.0020786 & 0.0020823 & 0.0020835  \\
				249 && $-0.0031016$ & $-0.0030747$ & $-0.0030745$ && 0.0016628 & 0.0016708 & 0.0016704  \\
				250 && $-0.0030911$ & $-0.0030623$ & $-0.0030621$ && 0.0016562 & 0.0016641 & 0.0016636  \\
				299 && $-0.0025435$ & $-0.0025586$ & $-0.0025562$ && 0.0013875 & 0.0013904 & 0.0013889  \\
				300 && $-0.0025436$ & $-0.0025500$ & $-0.0025476$ && 0.0013822 & 0.0013858 & 0.0013842  \\
				349 && $-0.0021615$ & $-0.0021908$ & $-0.0021869$ && 0.0012033 & 0.0011906 & 0.0011884  \\
				350 && $-0.0021702$ & $-0.0021846$ & $-0.0021806$ && 0.0012013 & 0.0011872 & 0.0011849  \\
				399 && $-0.0019214$ & $-0.0019155$ & $-0.0019106$ && 0.0010331 & 0.0010410 & 0.0010383  \\
				400 && $-0.0019273$ & $-0.0019107$ & $-0.0019058$ && 0.0010287 & 0.0010384 & 0.0010357  \\
				449 && $-0.0017199$ & $-0.0017017$ & $-0.0016960$ && 0.0009196 & 0.0009248 & 0.0009217  \\
				450 && $-0.0017210$ & $-0.0016979$ & $-0.0016922$ && 0.0009170 & 0.0009227 & 0.0009197  \\
				499 && $-0.0015134$ & $-0.0015308$ & $-0.0015247$ && 0.0008413 & 0.0008319 & 0.0008286  \\
				500 && $-0.0015061$ & $-0.0015277$ & $-0.0015216$ && 0.0008389 & 0.0008303 & 0.0008270  \\
				\hline
			\end{tabular}
		\end{small}
	\end{spacing}
\end{table}

	\clearpage
\begin{table}[htb]
	\caption{The values of $n\times\mathrm{Var}(\hat{\theta})$ for the median and the Hodges-Lehmann,
		and the values of $\mathrm{Var}(\hat{\theta})/(1-c_4(n)^2)$ for the MAD and Shamos
		($n=1,2,\ldots,50$).\label{TBL:nvar1}}
	\medskip
	\begin{spacing}{1.0}\centering
		\begin{small}
			\begin{tabular}{clcccclcc} 
				\hline
				$n$ && median & HL1    & HL2    &    HL3 && MAD    & Shamos  \\
				\cline{1-1} \cline{3-6} \cline{8-9}
				1 && 1.0000 &     NA & 1.0000 & 1.0000 &&     NA &     NA \\
				2 && 1.0000 & 1.0000 & 1.0000 & 1.0000 && 1.1000 & 2.2001 \\
				3 && 1.3463 & 1.0871 & 1.0221 & 1.0871 && 1.4372 & 2.3812 \\
				4 && 1.1930 & 1.0000 & 1.0949 & 1.0949 && 1.1680 & 1.6996 \\
				5 && 1.4339 & 1.0617 & 1.0754 & 1.0754 && 1.9809 & 1.8573 \\
				6 && 1.2882 & 1.0619 & 1.0759 & 1.0602 && 1.6859 & 1.7883 \\
				7 && 1.4736 & 1.0630 & 1.0814 & 1.0756 && 2.2125 & 1.6180 \\
				8 && 1.3459 & 1.0628 & 1.0728 & 1.0705 && 1.9486 & 1.5824 \\
				9 && 1.4957 & 1.0588 & 1.0756 & 1.0678 && 2.3326 & 1.5109 \\
				10 && 1.3833 & 1.0608 & 1.0743 & 1.0641 && 2.1072 & 1.4855 \\
				11 && 1.5088 & 1.0602 & 1.0693 & 1.0649 && 2.4082 & 1.4643 \\
				12 && 1.4087 & 1.0560 & 1.0670 & 1.0614 && 2.2112 & 1.4234 \\
				13 && 1.5195 & 1.0567 & 1.0685 & 1.0629 && 2.4570 & 1.4008 \\
				14 && 1.4298 & 1.0565 & 1.0663 & 1.0603 && 2.2848 & 1.3905 \\
				15 && 1.5249 & 1.0562 & 1.0645 & 1.0603 && 2.4952 & 1.3719 \\
				16 && 1.4457 & 1.0547 & 1.0637 & 1.0590 && 2.3412 & 1.3554 \\
				17 && 1.5302 & 1.0541 & 1.0633 & 1.0587 && 2.5217 & 1.3434 \\
				18 && 1.4585 & 1.0540 & 1.0621 & 1.0574 && 2.3846 & 1.3355 \\
				19 && 1.5333 & 1.0532 & 1.0605 & 1.0567 && 2.5447 & 1.3249 \\
				20 && 1.4702 & 1.0545 & 1.0620 & 1.0581 && 2.4185 & 1.3146 \\
				21 && 1.5383 & 1.0536 & 1.0611 & 1.0573 && 2.5611 & 1.3079 \\
				22 && 1.4770 & 1.0527 & 1.0596 & 1.0557 && 2.4475 & 1.3015 \\
				23 && 1.5420 & 1.0532 & 1.0597 & 1.0564 && 2.5758 & 1.2953 \\
				24 && 1.4850 & 1.0529 & 1.0594 & 1.0560 && 2.4699 & 1.2883 \\
				25 && 1.5438 & 1.0521 & 1.0586 & 1.0553 && 2.5873 & 1.2825 \\
				26 && 1.4896 & 1.0518 & 1.0578 & 1.0545 && 2.4886 & 1.2776 \\
				27 && 1.5462 & 1.0526 & 1.0582 & 1.0553 && 2.5960 & 1.2731 \\
				28 && 1.4954 & 1.0511 & 1.0567 & 1.0538 && 2.5030 & 1.2676 \\
				29 && 1.5476 & 1.0525 & 1.0581 & 1.0552 && 2.6070 & 1.2650 \\
				30 && 1.5005 & 1.0518 & 1.0571 & 1.0543 && 2.5199 & 1.2616 \\
				31 && 1.5482 & 1.0514 & 1.0564 & 1.0538 && 2.6132 & 1.2586 \\
				32 && 1.5057 & 1.0517 & 1.0567 & 1.0541 && 2.5335 & 1.2552 \\
				33 && 1.5516 & 1.0521 & 1.0571 & 1.0545 && 2.6208 & 1.2519 \\
				34 && 1.5091 & 1.0512 & 1.0560 & 1.0534 && 2.5442 & 1.2493 \\
				35 && 1.5515 & 1.0508 & 1.0554 & 1.0530 && 2.6285 & 1.2466 \\
				36 && 1.5123 & 1.0512 & 1.0557 & 1.0534 && 2.5545 & 1.2433 \\
				37 && 1.5531 & 1.0512 & 1.0556 & 1.0534 && 2.6332 & 1.2415 \\
				38 && 1.5148 & 1.0502 & 1.0545 & 1.0522 && 2.5637 & 1.2393 \\
				39 && 1.5550 & 1.0513 & 1.0555 & 1.0533 && 2.6344 & 1.2364 \\
				40 && 1.5173 & 1.0507 & 1.0547 & 1.0526 && 2.5720 & 1.2357 \\
				41 && 1.5532 & 1.0498 & 1.0539 & 1.0518 && 2.6403 & 1.2325 \\
				42 && 1.5206 & 1.0510 & 1.0549 & 1.0528 && 2.5780 & 1.2315 \\
				43 && 1.5552 & 1.0504 & 1.0541 & 1.0522 && 2.6436 & 1.2287 \\
				44 && 1.5224 & 1.0500 & 1.0537 & 1.0518 && 2.5869 & 1.2284 \\
				45 && 1.5568 & 1.0504 & 1.0541 & 1.0522 && 2.6477 & 1.2260 \\
				46 && 1.5240 & 1.0496 & 1.0533 & 1.0514 && 2.5904 & 1.2248 \\
				47 && 1.5570 & 1.0504 & 1.0539 & 1.0521 && 2.6511 & 1.2232 \\
				48 && 1.5249 & 1.0493 & 1.0528 & 1.0510 && 2.5960 & 1.2214 \\
				49 && 1.5562 & 1.0495 & 1.0529 & 1.0512 && 2.6537 & 1.2199 \\
				50 && 1.5267 & 1.0499 & 1.0532 & 1.0514 && 2.6014 & 1.2184 \\
				\hline
			\end{tabular}
		\end{small}
	\end{spacing}
\end{table}

\clearpage
\begin{table}[htb]
	\caption{The values of $n\times\mathrm{Var}(\hat{\theta})$ for the median and the Hodges-Lehmann,
		and the values of $\mathrm{Var}(\hat{\theta})/(1-c_4(n)^2)$ for the MAD and Shamos
		($n=51,52,\ldots,100$).\label{TBL:nvar2}}
	\medskip
	\begin{spacing}{1.0}\centering
		\begin{small}
			\begin{tabular}{clcccclcc} 
				\hline
				$n$ && median & HL1    & HL2    &    HL3 && MAD    & Shamos  \\
				\cline{1-1} \cline{3-6} \cline{8-9}
				51 && 1.5583 & 1.0502 & 1.0534 & 1.0517 && 2.6577 & 1.2199 \\
				52 && 1.5298 & 1.0499 & 1.0532 & 1.0515 && 2.6053 & 1.2174 \\
				53 && 1.5592 & 1.0501 & 1.0533 & 1.0517 && 2.6568 & 1.2160 \\
				54 && 1.5298 & 1.0489 & 1.0519 & 1.0503 && 2.6125 & 1.2156 \\
				55 && 1.5584 & 1.0493 & 1.0523 & 1.0508 && 2.6631 & 1.2144 \\
				56 && 1.5330 & 1.0497 & 1.0527 & 1.0512 && 2.6139 & 1.2132 \\
				57 && 1.5589 & 1.0496 & 1.0526 & 1.0510 && 2.6649 & 1.2126 \\
				58 && 1.5337 & 1.0495 & 1.0524 & 1.0509 && 2.6161 & 1.2098 \\
				59 && 1.5598 & 1.0501 & 1.0530 & 1.0515 && 2.6671 & 1.2095 \\
				60 && 1.5349 & 1.0489 & 1.0517 & 1.0503 && 2.6219 & 1.2095 \\
				61 && 1.5594 & 1.0492 & 1.0519 & 1.0505 && 2.6667 & 1.2073 \\
				62 && 1.5361 & 1.0492 & 1.0520 & 1.0505 && 2.6235 & 1.2071 \\
				63 && 1.5594 & 1.0485 & 1.0512 & 1.0498 && 2.6695 & 1.2064 \\
				64 && 1.5373 & 1.0494 & 1.0521 & 1.0507 && 2.6260 & 1.2050 \\
				65 && 1.5598 & 1.0488 & 1.0514 & 1.0500 && 2.6731 & 1.2067 \\
				66 && 1.5380 & 1.0496 & 1.0521 & 1.0508 && 2.6297 & 1.2036 \\
				67 && 1.5606 & 1.0494 & 1.0519 & 1.0506 && 2.6722 & 1.2034 \\
				68 && 1.5389 & 1.0491 & 1.0516 & 1.0503 && 2.6341 & 1.2030 \\
				69 && 1.5607 & 1.0479 & 1.0504 & 1.0491 && 2.6748 & 1.2025 \\
				70 && 1.5399 & 1.0490 & 1.0514 & 1.0502 && 2.6351 & 1.2016 \\
				71 && 1.5595 & 1.0482 & 1.0506 & 1.0494 && 2.6738 & 1.2005 \\
				72 && 1.5410 & 1.0491 & 1.0515 & 1.0503 && 2.6351 & 1.1993 \\
				73 && 1.5622 & 1.0492 & 1.0515 & 1.0503 && 2.6754 & 1.1993 \\
				74 && 1.5426 & 1.0498 & 1.0521 & 1.0510 && 2.6395 & 1.1990 \\
				75 && 1.5619 & 1.0489 & 1.0512 & 1.0500 && 2.6763 & 1.1985 \\
				76 && 1.5415 & 1.0486 & 1.0509 & 1.0497 && 2.6411 & 1.1975 \\
				77 && 1.5616 & 1.0485 & 1.0508 & 1.0496 && 2.6780 & 1.1975 \\
				78 && 1.5434 & 1.0494 & 1.0516 & 1.0505 && 2.6453 & 1.1971 \\
				79 && 1.5639 & 1.0493 & 1.0515 & 1.0504 && 2.6794 & 1.1968 \\
				80 && 1.5445 & 1.0497 & 1.0519 & 1.0508 && 2.6453 & 1.1958 \\
				81 && 1.5612 & 1.0486 & 1.0507 & 1.0496 && 2.6815 & 1.1960 \\
				82 && 1.5444 & 1.0494 & 1.0515 & 1.0504 && 2.6472 & 1.1947 \\
				83 && 1.5626 & 1.0484 & 1.0505 & 1.0494 && 2.6815 & 1.1947 \\
				84 && 1.5449 & 1.0490 & 1.0511 & 1.0500 && 2.6475 & 1.1939 \\
				85 && 1.5630 & 1.0484 & 1.0504 & 1.0494 && 2.6831 & 1.1938 \\
				86 && 1.5441 & 1.0479 & 1.0499 & 1.0489 && 2.6505 & 1.1931 \\
				87 && 1.5643 & 1.0495 & 1.0514 & 1.0504 && 2.6830 & 1.1923 \\
				88 && 1.5448 & 1.0478 & 1.0497 & 1.0487 && 2.6535 & 1.1929 \\
				89 && 1.5640 & 1.0487 & 1.0506 & 1.0496 && 2.6857 & 1.1931 \\
				90 && 1.5463 & 1.0483 & 1.0503 & 1.0493 && 2.6562 & 1.1920 \\
				91 && 1.5634 & 1.0486 & 1.0505 & 1.0495 && 2.6853 & 1.1914 \\
				92 && 1.5477 & 1.0491 & 1.0509 & 1.0500 && 2.6567 & 1.1913 \\
				93 && 1.5631 & 1.0481 & 1.0500 & 1.0490 && 2.6859 & 1.1906 \\
				94 && 1.5482 & 1.0488 & 1.0507 & 1.0497 && 2.6584 & 1.1907 \\
				95 && 1.5629 & 1.0481 & 1.0499 & 1.0490 && 2.6878 & 1.1905 \\
				96 && 1.5466 & 1.0477 & 1.0495 & 1.0486 && 2.6576 & 1.1894 \\
				97 && 1.5636 & 1.0480 & 1.0498 & 1.0489 && 2.6881 & 1.1895 \\
				98 && 1.5477 & 1.0477 & 1.0495 & 1.0486 && 2.6613 & 1.1899 \\
				99 && 1.5642 & 1.0483 & 1.0501 & 1.0492 && 2.6888 & 1.1887 \\
				100 && 1.5484 & 1.0481 & 1.0498 & 1.0489 && 2.6604 & 1.1874 \\
				\hline
			\end{tabular}
		\end{small}
	\end{spacing}
\end{table}

\clearpage
\begin{table}[htb]
	\caption{The values of $n\times\mathrm{Var}(\hat{\theta})$ for the median and the Hodges-Lehmann,
		and the values of $\mathrm{Var}(\hat{\theta})/(1-c_4(n)^2)$ for the MAD and Shamos
		($n=109,110, 119,120, \ldots, 499,500$).\label{TBL:nvar3}}
	\medskip
	\begin{spacing}{1.0}\centering
		\begin{small}
			\begin{tabular}{clcccclcc} 
				\hline
				$n$ && median & HL1    & HL2    &    HL3 && MAD    & Shamos  \\
				\cline{1-1} \cline{3-6} \cline{8-9}
				109 && 1.5655 & 1.0490 & 1.0506 & 1.0498 && 2.6889 & 1.1857 \\
				110 && 1.5508 & 1.0484 & 1.0500 & 1.0492 && 2.6657 & 1.1856 \\
				119 && 1.5651 & 1.0478 & 1.0492 & 1.0485 && 2.6936 & 1.1830 \\
				120 && 1.5526 & 1.0478 & 1.0493 & 1.0486 && 2.6717 & 1.1836 \\
				129 && 1.5661 & 1.0477 & 1.0490 & 1.0483 && 2.6953 & 1.1809 \\
				130 && 1.5541 & 1.0478 & 1.0492 & 1.0485 && 2.6727 & 1.1804 \\
				139 && 1.5671 & 1.0491 & 1.0503 & 1.0497 && 2.6963 & 1.1792 \\
				140 && 1.5567 & 1.0495 & 1.0508 & 1.0502 && 2.6770 & 1.1794 \\
				149 && 1.5666 & 1.0484 & 1.0496 & 1.0490 && 2.7008 & 1.1789 \\
				150 && 1.5566 & 1.0484 & 1.0496 & 1.0490 && 2.6815 & 1.1788 \\
				159 && 1.5673 & 1.0484 & 1.0495 & 1.0490 && 2.7006 & 1.1768 \\
				160 && 1.5584 & 1.0485 & 1.0495 & 1.0490 && 2.6827 & 1.1765 \\
				169 && 1.5661 & 1.0474 & 1.0485 & 1.0479 && 2.7012 & 1.1757 \\
				170 && 1.5578 & 1.0476 & 1.0486 & 1.0481 && 2.6861 & 1.1755 \\
				179 && 1.5676 & 1.0477 & 1.0487 & 1.0482 && 2.7038 & 1.1750 \\
				180 && 1.5590 & 1.0480 & 1.0490 & 1.0485 && 2.6889 & 1.1750 \\
				189 && 1.5663 & 1.0473 & 1.0483 & 1.0478 && 2.7043 & 1.1743 \\
				190 && 1.5584 & 1.0473 & 1.0482 & 1.0478 && 2.6903 & 1.1741 \\
				199 && 1.5681 & 1.0481 & 1.0490 & 1.0486 && 2.7049 & 1.1741 \\
				200 && 1.5608 & 1.0482 & 1.0491 & 1.0486 && 2.6904 & 1.1732 \\
				249 && 1.5679 & 1.0472 & 1.0479 & 1.0476 && 2.7083 & 1.1705 \\
				250 && 1.5623 & 1.0479 & 1.0486 & 1.0483 && 2.6977 & 1.1709 \\
				299 && 1.5689 & 1.0477 & 1.0483 & 1.0480 && 2.7084 & 1.1673 \\
				300 && 1.5642 & 1.0479 & 1.0485 & 1.0482 && 2.6986 & 1.1670 \\
				349 && 1.5700 & 1.0479 & 1.0484 & 1.0481 && 2.7131 & 1.1673 \\
				350 && 1.5654 & 1.0479 & 1.0484 & 1.0481 && 2.7049 & 1.1675 \\
				399 && 1.5691 & 1.0475 & 1.0479 & 1.0477 && 2.7126 & 1.1650 \\
				400 && 1.5646 & 1.0469 & 1.0474 & 1.0472 && 2.7072 & 1.1651 \\
				449 && 1.5694 & 1.0474 & 1.0478 & 1.0476 && 2.7125 & 1.1639 \\
				450 && 1.5659 & 1.0475 & 1.0479 & 1.0477 && 2.7056 & 1.1645 \\
				499 && 1.5701 & 1.0475 & 1.0479 & 1.0477 && 2.7147 & 1.1637 \\
				500 && 1.5674 & 1.0482 & 1.0486 & 1.0484 && 2.7101 & 1.1646 \\
				\hline
			\end{tabular}
		\end{small}
	\end{spacing}
\end{table}

\clearpage
\begin{table}[htb]
	\caption{Relative efficiencies of the median and Hodges-Lehmann
		estimators to the sample mean and those of the MAD and Shamos estimators
		to the sample standard deviation under the normal distribution ($n=1,2,\ldots,50$).\label{TBL:RE1}}
	\medskip
	\begin{spacing}{1.0}\centering
		\begin{small}
			\begin{tabular}{clcccclcc} 
				\hline
				$n$ && median & HL1    & HL2    &    HL3 && MAD    & Shamos  \\
				\cline{1-1} \cline{3-6} \cline{8-9}
				1 && 1.0000 & NA     & 1.0000 & 1.0000 && NA     & NA      \\
				2 && 1.0000 & 1.0000 & 1.0000 & 1.0000 && 0.9091 & 0.4545  \\
				3 && 0.7427 & 0.9199 & 0.9784 & 0.9199 && 0.6958 & 0.4199  \\
				4 && 0.8382 & 1.0000 & 0.9133 & 0.9133 && 0.8562 & 0.5884  \\
				5 && 0.6974 & 0.9419 & 0.9299 & 0.9299 && 0.5048 & 0.5384  \\
				6 && 0.7763 & 0.9417 & 0.9295 & 0.9432 && 0.5932 & 0.5592  \\
				7 && 0.6786 & 0.9407 & 0.9248 & 0.9297 && 0.4520 & 0.6180  \\
				8 && 0.7430 & 0.9409 & 0.9322 & 0.9342 && 0.5132 & 0.6320  \\
				9 && 0.6686 & 0.9445 & 0.9297 & 0.9365 && 0.4287 & 0.6618  \\
				10 && 0.7229 & 0.9426 & 0.9308 & 0.9398 && 0.4746 & 0.6732  \\
				11 && 0.6628 & 0.9432 & 0.9352 & 0.9391 && 0.4153 & 0.6829  \\
				12 && 0.7098 & 0.9470 & 0.9372 & 0.9422 && 0.4522 & 0.7026  \\
				13 && 0.6581 & 0.9464 & 0.9359 & 0.9408 && 0.4070 & 0.7139  \\
				14 && 0.6994 & 0.9465 & 0.9378 & 0.9432 && 0.4377 & 0.7192  \\
				15 && 0.6558 & 0.9468 & 0.9394 & 0.9432 && 0.4008 & 0.7289  \\
				16 && 0.6917 & 0.9482 & 0.9402 & 0.9443 && 0.4271 & 0.7378  \\
				17 && 0.6535 & 0.9486 & 0.9405 & 0.9445 && 0.3966 & 0.7444  \\
				18 && 0.6856 & 0.9487 & 0.9415 & 0.9457 && 0.4194 & 0.7488  \\
				19 && 0.6522 & 0.9495 & 0.9430 & 0.9463 && 0.3930 & 0.7547  \\
				20 && 0.6802 & 0.9483 & 0.9416 & 0.9451 && 0.4135 & 0.7607  \\
				21 && 0.6501 & 0.9491 & 0.9424 & 0.9458 && 0.3905 & 0.7646  \\
				22 && 0.6770 & 0.9499 & 0.9437 & 0.9472 && 0.4086 & 0.7684  \\
				23 && 0.6485 & 0.9495 & 0.9437 & 0.9466 && 0.3882 & 0.7720  \\
				24 && 0.6734 & 0.9498 & 0.9440 & 0.9470 && 0.4049 & 0.7762  \\
				25 && 0.6478 & 0.9504 & 0.9447 & 0.9476 && 0.3865 & 0.7798  \\
				26 && 0.6713 & 0.9507 & 0.9454 & 0.9483 && 0.4018 & 0.7827  \\
				27 && 0.6468 & 0.9501 & 0.9450 & 0.9476 && 0.3852 & 0.7855  \\
				28 && 0.6687 & 0.9514 & 0.9463 & 0.9490 && 0.3995 & 0.7889  \\
				29 && 0.6462 & 0.9501 & 0.9451 & 0.9477 && 0.3836 & 0.7905  \\
				30 && 0.6664 & 0.9507 & 0.9459 & 0.9485 && 0.3968 & 0.7926  \\
				31 && 0.6459 & 0.9511 & 0.9466 & 0.9489 && 0.3827 & 0.7945  \\
				32 && 0.6641 & 0.9508 & 0.9463 & 0.9486 && 0.3947 & 0.7967  \\
				33 && 0.6445 & 0.9505 & 0.9460 & 0.9483 && 0.3816 & 0.7988  \\
				34 && 0.6627 & 0.9513 & 0.9470 & 0.9493 && 0.3931 & 0.8004  \\
				35 && 0.6445 & 0.9516 & 0.9475 & 0.9496 && 0.3804 & 0.8022  \\
				36 && 0.6612 & 0.9513 & 0.9472 & 0.9493 && 0.3915 & 0.8043  \\
				37 && 0.6439 & 0.9513 & 0.9473 & 0.9493 && 0.3798 & 0.8055  \\
				38 && 0.6601 & 0.9522 & 0.9483 & 0.9504 && 0.3901 & 0.8069  \\
				39 && 0.6431 & 0.9512 & 0.9475 & 0.9494 && 0.3796 & 0.8088  \\
				40 && 0.6591 & 0.9518 & 0.9481 & 0.9500 && 0.3888 & 0.8093  \\
				41 && 0.6438 & 0.9525 & 0.9489 & 0.9507 && 0.3787 & 0.8113  \\
				42 && 0.6576 & 0.9515 & 0.9479 & 0.9498 && 0.3879 & 0.8120  \\
				43 && 0.6430 & 0.9521 & 0.9486 & 0.9504 && 0.3783 & 0.8138  \\
				44 && 0.6569 & 0.9524 & 0.9490 & 0.9508 && 0.3866 & 0.8141  \\
				45 && 0.6423 & 0.9520 & 0.9487 & 0.9504 && 0.3777 & 0.8157  \\
				46 && 0.6562 & 0.9527 & 0.9494 & 0.9511 && 0.3860 & 0.8164  \\
				47 && 0.6422 & 0.9520 & 0.9489 & 0.9505 && 0.3772 & 0.8175  \\
				48 && 0.6558 & 0.9530 & 0.9499 & 0.9515 && 0.3852 & 0.8187  \\
				49 && 0.6426 & 0.9529 & 0.9497 & 0.9513 && 0.3768 & 0.8197  \\
				50 && 0.6550 & 0.9525 & 0.9495 & 0.9511 && 0.3844 & 0.8208  \\
				\hline
			\end{tabular}
		\end{small}
	\end{spacing}
\end{table}

\clearpage
\begin{table}[htb]
	\caption{Relative efficiencies of the median and Hodges-Lehmann
		estimators to the sample mean and those of the MAD and Shamos estimators
		to the sample standard deviation under the normal distribution ($n=51,52,\ldots,100$).\label{TBL:RE2}}
	\medskip
	\begin{spacing}{1.0}\centering
		\begin{small}
			\begin{tabular}{clcccclcc} 
				\hline
				$n$ && median & HL1    & HL2    &    HL3 && MAD    & Shamos  \\
				\cline{1-1} \cline{3-6} \cline{8-9}
				51 && 0.6417 & 0.9522 & 0.9493 & 0.9508 && 0.3763 & 0.8197  \\
				52 && 0.6537 & 0.9525 & 0.9495 & 0.9510 && 0.3838 & 0.8214  \\
				53 && 0.6414 & 0.9523 & 0.9494 & 0.9509 && 0.3764 & 0.8223  \\
				54 && 0.6537 & 0.9534 & 0.9506 & 0.9521 && 0.3828 & 0.8226  \\
				55 && 0.6417 & 0.9530 & 0.9503 & 0.9517 && 0.3755 & 0.8234  \\
				56 && 0.6523 & 0.9527 & 0.9499 & 0.9513 && 0.3826 & 0.8243  \\
				57 && 0.6415 & 0.9528 & 0.9501 & 0.9514 && 0.3752 & 0.8247  \\
				58 && 0.6520 & 0.9528 & 0.9502 & 0.9515 && 0.3822 & 0.8266  \\
				59 && 0.6411 & 0.9523 & 0.9497 & 0.9510 && 0.3749 & 0.8268  \\
				60 && 0.6515 & 0.9534 & 0.9508 & 0.9521 && 0.3814 & 0.8268  \\
				61 && 0.6413 & 0.9531 & 0.9506 & 0.9519 && 0.3750 & 0.8283  \\
				62 && 0.6510 & 0.9531 & 0.9506 & 0.9519 && 0.3812 & 0.8284  \\
				63 && 0.6413 & 0.9538 & 0.9513 & 0.9526 && 0.3746 & 0.8289  \\
				64 && 0.6505 & 0.9529 & 0.9505 & 0.9517 && 0.3808 & 0.8299  \\
				65 && 0.6411 & 0.9535 & 0.9511 & 0.9523 && 0.3741 & 0.8287  \\
				66 && 0.6502 & 0.9528 & 0.9505 & 0.9516 && 0.3803 & 0.8308  \\
				67 && 0.6408 & 0.9529 & 0.9507 & 0.9518 && 0.3742 & 0.8310  \\
				68 && 0.6498 & 0.9532 & 0.9509 & 0.9521 && 0.3796 & 0.8312  \\
				69 && 0.6408 & 0.9543 & 0.9520 & 0.9532 && 0.3739 & 0.8316  \\
				70 && 0.6494 & 0.9533 & 0.9511 & 0.9522 && 0.3795 & 0.8323  \\
				71 && 0.6412 & 0.9540 & 0.9518 & 0.9529 && 0.3740 & 0.8330  \\
				72 && 0.6489 & 0.9532 & 0.9510 & 0.9521 && 0.3795 & 0.8338  \\
				73 && 0.6401 & 0.9532 & 0.9510 & 0.9521 && 0.3738 & 0.8338  \\
				74 && 0.6483 & 0.9525 & 0.9504 & 0.9515 && 0.3789 & 0.8341  \\
				75 && 0.6402 & 0.9534 & 0.9513 & 0.9524 && 0.3736 & 0.8344  \\
				76 && 0.6487 & 0.9536 & 0.9516 & 0.9526 && 0.3786 & 0.8351  \\
				77 && 0.6404 & 0.9537 & 0.9517 & 0.9527 && 0.3734 & 0.8351  \\
				78 && 0.6479 & 0.9529 & 0.9509 & 0.9520 && 0.3780 & 0.8353  \\
				79 && 0.6394 & 0.9530 & 0.9510 & 0.9520 && 0.3732 & 0.8355  \\
				80 && 0.6475 & 0.9526 & 0.9507 & 0.9517 && 0.3780 & 0.8363  \\
				81 && 0.6405 & 0.9537 & 0.9517 & 0.9527 && 0.3729 & 0.8361  \\
				82 && 0.6475 & 0.9529 & 0.9510 & 0.9520 && 0.3778 & 0.8370  \\
				83 && 0.6400 & 0.9538 & 0.9520 & 0.9529 && 0.3729 & 0.8370  \\
				84 && 0.6473 & 0.9533 & 0.9514 & 0.9523 && 0.3777 & 0.8376  \\
				85 && 0.6398 & 0.9539 & 0.9520 & 0.9529 && 0.3727 & 0.8376  \\
				86 && 0.6476 & 0.9543 & 0.9525 & 0.9534 && 0.3773 & 0.8381  \\
				87 && 0.6393 & 0.9529 & 0.9511 & 0.9520 && 0.3727 & 0.8387  \\
				88 && 0.6473 & 0.9544 & 0.9526 & 0.9535 && 0.3769 & 0.8383  \\
				89 && 0.6394 & 0.9536 & 0.9518 & 0.9527 && 0.3723 & 0.8382  \\
				90 && 0.6467 & 0.9539 & 0.9521 & 0.9530 && 0.3765 & 0.8389  \\
				91 && 0.6396 & 0.9536 & 0.9519 & 0.9528 && 0.3724 & 0.8394  \\
				92 && 0.6461 & 0.9532 & 0.9515 & 0.9524 && 0.3764 & 0.8394  \\
				93 && 0.6398 & 0.9541 & 0.9524 & 0.9533 && 0.3723 & 0.8399  \\
				94 && 0.6459 & 0.9535 & 0.9518 & 0.9526 && 0.3762 & 0.8399  \\
				95 && 0.6398 & 0.9541 & 0.9524 & 0.9533 && 0.3721 & 0.8400  \\
				96 && 0.6466 & 0.9545 & 0.9529 & 0.9537 && 0.3763 & 0.8407  \\
				97 && 0.6396 & 0.9542 & 0.9526 & 0.9534 && 0.3720 & 0.8407  \\
				98 && 0.6461 & 0.9545 & 0.9529 & 0.9537 && 0.3757 & 0.8404  \\
				99 && 0.6393 & 0.9539 & 0.9523 & 0.9531 && 0.3719 & 0.8412  \\
				100 && 0.6458 & 0.9541 & 0.9525 & 0.9533 && 0.3759 & 0.8422  \\
				\hline
			\end{tabular}
		\end{small}
	\end{spacing}
\end{table}

	
\begin{figure}[t]
	\centering
	\includegraphics{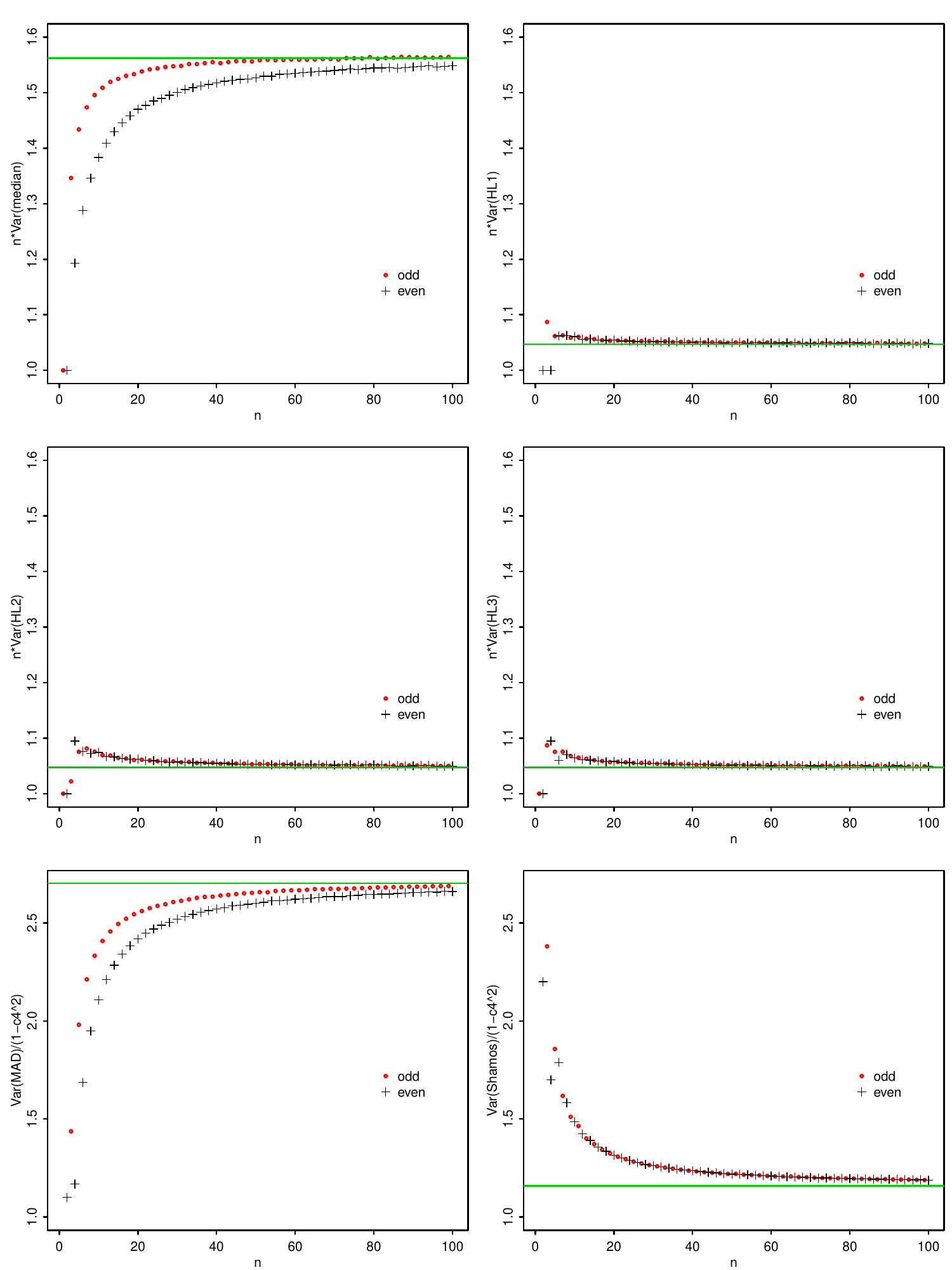}
	\caption{The values of $n\,\mathrm{Var}(\mathrm{median})$,
		$n\,\mathrm{Var}(\mathrm{HL})$,
		${\mathrm{Var}(\mathrm{MAD})}/{(1-c_4(n)^2)}$, and
		${\mathrm{Var}(\mathrm{Shamos})}/{(1-c_4(n)^2)}$. \label{FIG:nvar}}
\end{figure}

\begin{figure}[t]
	\centering
	\includegraphics{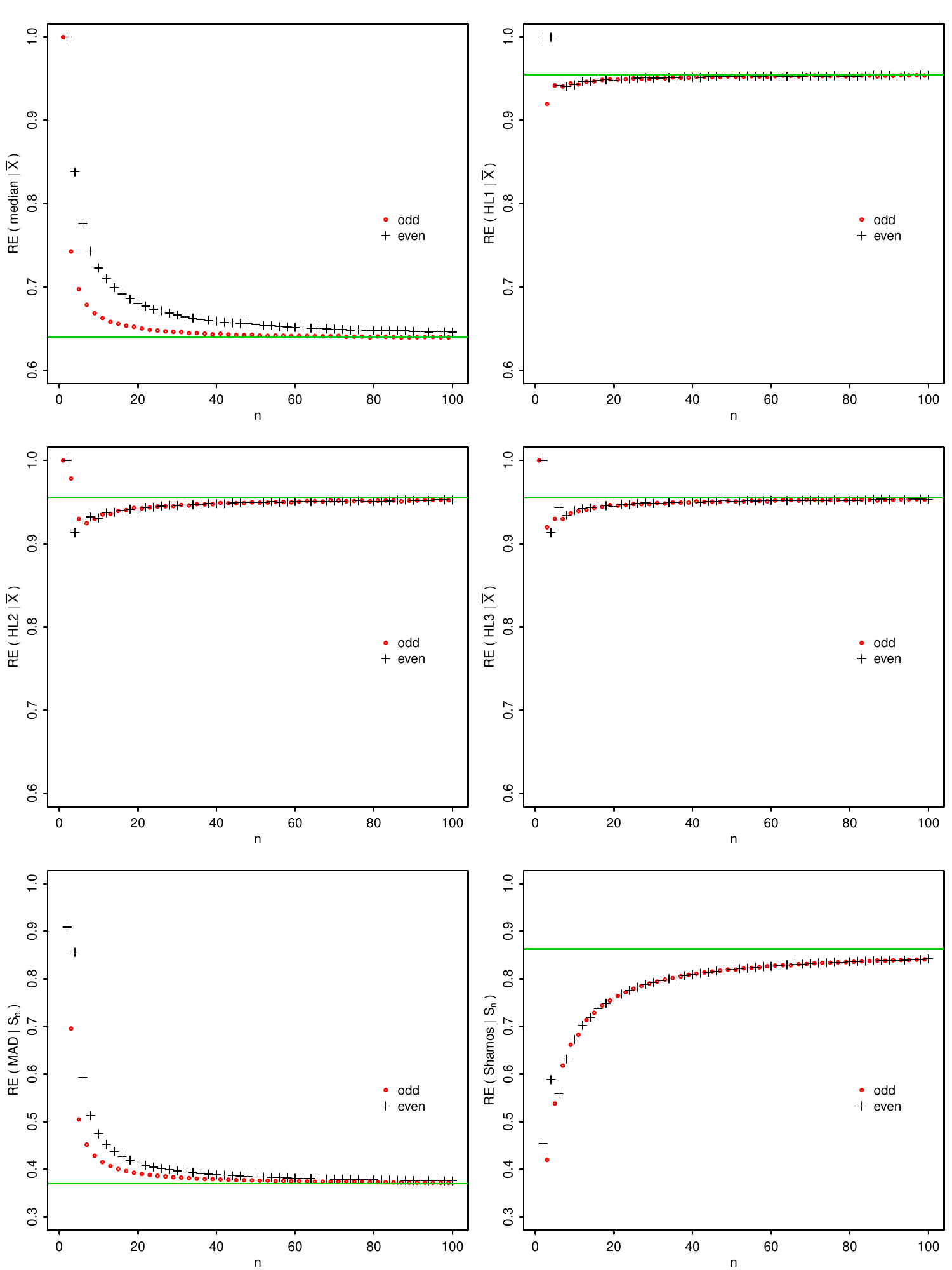}
	\caption{The relative efficiencies under consideration.\label{FIG:RE}}
\end{figure}

\end{document}